\documentclass[aps,prd,showpacs,onecolumn,preprintnumbers,notitlepage,superscriptaddress,
nofootinbib,amsfont,amssymb,10pt,singlespacing] {revtex4-1}
\usepackage{amsmath,bm}
\usepackage{times}
\usepackage{braket}
\usepackage{color,graphicx}
\usepackage{slashed}
\usepackage{hyperref}

\newcommand{\beq}{\begin{eqnarray}}
\newcommand{\eeq}{\end{eqnarray}}
\newcommand{\non}{\nonumber\\ }
\newcommand{\ov}{\overline}

\newcommand{\acp}{A_{CP}}
\newcommand{\psl}{ P \hspace{-2.8truemm}/ }
\newcommand{\nsl}{ n \hspace{-2.2truemm}/ }
\newcommand{\vsl}{ v \hspace{-2.2truemm}/ }
\newcommand{\epsl}{\epsilon \hspace{-1.8truemm}/\,  }
\newcommand{\kst} {K_0^{*}(1430)}

\newcommand{\ks} {K_0^*}
\newcommand{\ksb} {\overline{K}_0^*}

\def\lsim{ {\ \lower-1.2pt\vbox{\hbox{\rlap{$<$}\lower6pt\vbox{\hbox{$\sim$}
}}}\ } }
\def\gsim{ {\ \lower-1.2pt\vbox{\hbox{\rlap{$>$}\lower6pt\vbox{\hbox{$\sim$}
}}}\ } }
\def \epjc{ Eur. Phys. J. C }
\def \epl{ Eur. Phys. Lett. }
\def \jpg{  J. Phys. G }
\def \npb{  Nucl. Phys. B }
\def \npps { Nucl. Phys. B (Proc. Suppl.) }
\def \plb{  Phys. Lett. B }

\def \prd{  Phys. Rev. D }
\def \prl{  Phys. Rev. Lett.  }

\def \jhep{ J. High Energy Phys.  }

\def \cpc{ Chin. Phys. C }
\def \ctp{ Commun. Theor. Phys. }
\def \rmp{ Rev. Mod. Phys. }
\def \ppnp{ Prog. Part. $\&$ Nucl. Phys. }
\def \arnps{ Ann. Rev. Nucl. Part. Sci. }

\definecolor{Red}{rgb}{1.,0.,0.}

\definecolor{Blue}{rgb}{0.,0.,1.}

\definecolor{nicered}{rgb}{0.7,0.1,0.1}
\definecolor{nicegreen}{rgb}{0.1,0.5,0.1}

\hypersetup{colorlinks,citecolor=nicegreen,linkcolor=nicered}

\begin{document}

\title{ Branching Ratios and CP Violations of $B \to K_0^*(1430) K^*$
Decays in the pQCD Approach}

\author{Xin~Liu}
\email[Electronic address:]{liuxin.physics@gmail.com}
\affiliation{School of Physics and Electronic Engineering,\\ Jiangsu Normal University, Xuzhou, Jiangsu 221116,
People's Republic of China}

\author{Zhen-Jun~Xiao}
\email[Electronic address:]{xiaozhenjun@njnu.edu.cn}
\affiliation{Department of Physics and Institute of Theoretical
Physics,\\ Nanjing Normal University, Nanjing, Jiangsu 210023,
People's Republic of China}

\author{Zhi-Tian~Zou}
\email[Electronic address:]{zouzt@ihep.ac.cn}
\affiliation{Department of Physics, Yantai University, Yantai, Shandong 264005, People's Republic of China}

\date{\today}

\begin{abstract}

We investigate $B \to K_0^*(1430) K^*$ decays in the perturbative QCD(pQCD)
factorization approach, where $B$ denotes $B_u$, $B_d$ and $B_s$ meson respectively,
and the scalar $K_0^*(1430)$ is considered as a meson based on
the model of conventional two-quark structure.
With the light-cone distribution amplitude of $K_0^*(1430)$ defined in two scenarios, namely Scenario 1 and
Scenario 2, we make the first estimation for the branching ratios and
CP-violating asymmetries for those concerned decay modes in the pQCD factorization approach.
For all considered $B \to \kst K^*$ decays in this paper,  only one preliminary
upper limit on the branching ratio of $B^0 \to {\kst}^0 \overline{K}^{*0}$ measured at 90\% C.L.
by Belle Collaboration is available now.
It is therefore of great interest to examine the predicted physical quantities
at two $B$ factories, Large Hadron Collider experiments, and forthcoming
Super-$B$ facility, then test the reliability of the pQCD approach employed
to study the considered decay modes involving a $p$-wave scalar meson as one of the final state mesons.
Furthermore, these pQCD predictions combined with the future precision measurements are also
helpful to explore the complicated QCD dynamics involved in the light scalars.

\end{abstract}

\preprint{\footnotesize JSNU/PHY-TH-2013}
\maketitle

%
%

\section{Introduction}

The inner structure of the light scalars, generally below 2 GeV, has been explored
by the physicists at both experimental and theoretical aspects for several decades.
However, unfortunately, their underlying structure has not yet been well established
and the identification of the considered light scalars is known as a long-standing puzzle
~\cite{Beringer12:pdg}.
But, it is lucky for us that the light scalars could be studied in the decay channels
of heavy flavor $B$ mesons, with the rich data provided by $B$ factories,
the Large Hadron Collider(LHC) experiments ~\cite{Cheng09:bdecays}, and the forthcoming
Super-$B$ factory~\cite{Soni07,SuperB}.
Ever since the $B \to f_0(980) K$ mode was firstly measured by
Belle Collaboration in 2002~\cite{belle02:f98}, then confirmed by BaBar
Collaboration in 2004~\cite{babar04:f98},
more and more channels with $p$-wave light scalars in the final states
of $B$ meson decays have been opened and more precise data have been obtained
~\cite{Beringer12:pdg}.
With the gradually enlarging data samples collected in the  running LHC experiments
and the forthcoming Super-$B$ factory,
it is therefore believed with enough reasons that as a different unique
insight to the nature of the light scalars, the $B$ meson decays
involving $p$-wave light scalars will provide good places in investigating
the physical properties of light scalars.
It is expected that the old puzzles related to the nature of the scalars could receive
new attention through the studies on rare $B$ meson decays involving scalars,
apart from those well-known primary tasks in heavy flavor physics.

Although the underlying structure of the light scalars is still controversial,
the scalar $a_0(1450)$ has been confirmed to be a conventional $q\bar{q}$ meson in lattice
calculations~\cite{Mathur06,Kim97,Bardeen02,Kunihiro04:lattice,Prelovsek04:lattice}
recently. Furthermore, a good SU(3) flavor symmetry is indicated in
the scalar sector through the calculations in lattice QCD~\cite{Mathur06}
on the masses of $a_0(1450)$ and $K_0^*(1430)$. The evaluations on the relevant
$K_0^*(1430)$ (Hereafter, unless otherwise stated,
$K_0^*$ will be adopted to describe the $K_0^*(1430)$ throughout
the paper for the sake of simplicity.) modes therefore draw more attention now.
Recently, the authors in Ref.~\cite{Cheng06:B2SP} proposed two possible
scenarios, namely, Scenario 1(S1) and Scenario 2(S2), to describe the components
of $\ks$ meson in the QCD sum rule method based on the assumption of
conventional two-quark structure:
\begin{itemize}
\item
In S1, the lighter state $\kappa$ near 1 GeV is treated as the lowest lying $q\bar q$
state, while the heavier state $\ks$ above 1 GeV is considered as the corresponding first excited $q \bar q$ state.
\item
In S2, $\ks$ is regarded as ground $q\bar q$ state and the corresponding
first excited state lies between (2.0 $\sim$ 2.3) GeV. Then $\kappa$ is viewed
as the four-quark bound state or hybrid state.
\end{itemize}

The two body charmless hadronic $B$ meson decays to the scalar $\ks$ meson
have been studied intensively, for example,
in Refs.~\cite{Delepine08:b2sp,Cheng06:B2SP,Zhang:scalar,Liu:scalar,Wang:scalar} by employing
different factorization approaches respectively, or in Ref.~\cite{Li:scalar}
even with the inclusion of the new physics contributions from a $Z'$ boson.
This year, the authors of Ref.~\cite{Cheng13:scalar} revisited the $B \to SP, SV$
decays in  the framework of QCD factorization.

On the theory side, it is necessary for us to make all possible
investigations on the decay modes of $B$ meson with the scalar $\ks$ to identify the
favorite one from the proposed S1 and S2 scenarios, which will also be helpful
to obtain the new insights in the properties of the scalar $\ks$;
On the experiment side, however, so far only a preliminary
upper limit at 90\% C.L. on the branching ratio of $B^0 \to {\ks}^0 \ov{K}^{*0}$
decay has been measured by Belle Collaboration~\cite{Chiang10:k0k0b},
\beq
  Br(B^0 \to {\ks}^0 \ov{K}^{*0}) <  3.3 \times 10^{-6} \;.
  \label{eq:exp1}
\eeq
Of course, this measurement would be improved rapidly with the LHC experiments at CERN,
and other relevant channels considered in this work would also be observed
in the near future.

In this work, we will study the branching ratios and CP-violating asymmetries
of $B \to K_0^* K^*$ decays in the standard model (SM) by employing the low energy
effective Hamiltonian~\cite{Buras96:weak} and the
pQCD factorization approach~\cite{Keum01:kpi,Lu01:pipi,Li03:ppnp},
where $B$ stands for $B_{u,d}$ and $B_s$ respectively.
Based on $k_T$ factorization, the pQCD approach is one of the popular factorization
methods for dealing with the B meson exclusive decays.
In the pQCD approach, the parton transverse momentum $k_T$ is kept in order to
eliminate the end-point singularity, while the Sudakov factor
play an important role in suppressing the long-distance contribution \cite{Li03:ppnp}.
We here not only consider the usual factorizable emission diagrams, but also evaluate the nonfactorizable spectator
and the annihilation type contributions simultaneously.
As far as the annihilation contributions are concerned, both the soft-collinear effective theory~\cite{scet:factor}
and the pQCD approach can work, but with rather different viewpoints on the
relevant perturbative calculations~\cite{Stewart08:anni-scet,Chay08:complexanni}.
However, the predictions on the pure annihilation decays based on the pQCD approach
can accommodate the experimental data well, for example,
for the $B_s\to \pi^+\pi^-$ and $B^0\to K^+K^-$ decays as have been done
in Refs.~\cite{Lu03:dsk,Li04:pippim,Lu07:bs2mm,Xiao11:pippim}.
In this work, we will therefore leave the controversies aside and adopt this approach in our analysis.

The paper is organized as follows. Section~\ref{sec:form} is devoted to the ingredients of the basic formalism
in the pQCD approach. The analytic expressions for the decay amplitudes of $B \to \ks K^*$ modes
in the pQCD approach are also collected in this section. The numerical results and phenomenological analysis
for the branching ratios and CP-violating asymmetries of the considered decays
are given in Sec.~\ref{sec:randd}. We summarize and conclude in Sec.~\ref{sec:summary}.

%
%
\section{ Formalism}\label{sec:form}

The pQCD approach is one of the popular methods to evaluate the hadronic matrix elements in the heavy $b$-flavor
mesons' decays. The basic idea of the pQCD approach is that it takes into
account the transverse momentum $k_T$ of the valence quarks
in the calculation of the hadronic matrix elements.
The $B$ meson transition form factors, and the spectator and
annihilation contributions are then all calculable in the framework
of the $k_T$ factorization, where three energy scales $m_W,m_B$ and $t\approx \sqrt{m_B \Lambda_{\rm QCD}} $
are involved~\cite{Keum01:kpi,Lu01:pipi,Li95-96:pQCD}.
The running of the Wilson coefficients $C_i(t)$ with $t \geq \sqrt{m_B \Lambda_{\rm QCD}} $ are controlled by
the renormalization group equation (RGE) and can be  calculated perturbatively.
The dynamics below $\sqrt{m_B \Lambda_{\rm QCD}}$ is soft, which is described by the meson wave
functions. The soft dynamics is not perturbative but universal for all channels.
In the pQCD approach, a $B \to M_2 M_3$ decay amplitude is therefore factorized into the convolution of the six-quark
hard kernel($H$), the jet function($J$) and the Sudakov factor($S$) with the
bound-state wave functions($\Phi$) as follows,
\begin{eqnarray}
{\cal A}(B \to M_2 M_3)=\Phi_{B} \otimes H \otimes J \otimes S
\otimes \Phi_{M_2} \otimes \Phi_{M_3}\;, \label{eq:sixquarks}
\end{eqnarray}
The jet function $J$ comes from the threshold resummation, which
exhibits strong suppression effect in the small $x$ (quark momentum fraction) region~\cite{Li02:threshold}.
The Sudakov factor $S$ comes from the $k_T$ resummation, which
provide a strong suppression in the small $k_T$ region~\cite{Sterman89-92:ktfact}.
Therefore, these resummation effects guarantee the removal of the endpoint singularities.

\subsection{Wave Functions and Distribution Amplitudes}\label{ssec:wf}

Throughout this paper, we will use light-cone coordinate $(P^+, P^-, {\bf P_T})$ to describe the meson's momenta with the definitions
$P^{\pm}=(p_0 \pm p_3)/\sqrt{2}$ and ${\bf P_T}=(p_1,p_2)$.
The heavy $B$ meson is usually treated as a heavy-light system and its light-cone wave function
can generally be defined as~\cite{Lu01:pipi,Keum01:kpi,Lu03:form}
\beq
\Phi_{B,\alpha\beta,ij}&\equiv&
  \langle 0|\bar{b}_{\beta j}(0)q_{\alpha i}(z)|B(P)\rangle \non
&=& \frac{i \delta_{ij}}{\sqrt{2N_c}}\int dx d^2 k_T e^{-i (xP^-z^+ - k_T z_T)}
\left\{(\psl +m_{B})\gamma_5
 \phi_{B}(x, k_T) \right\}_{\alpha\beta}\;;
\label{eq:def-bq}
\eeq
where the indices $i,j$ and $\alpha,\beta$ are the Lorentz indices and color indices, respectively,
$P(m)$ is the momentum(mass) of the $B$ meson, $N_c$ is the color factor, and
$k_T$ is the intrinsic transverse momentum
of the light quark in $B$ meson. Note that, in principle, there are two Lorentz
structures of the wave function to be considered in the numerical calculations, however,
the contribution induced by the second Lorentz structure
is numerically small and approximately negligible~\cite{Lu03:form}.

In Eq.~(\ref{eq:def-bq}),
$\phi_{B}(x,k_T)$ is the $B$ meson distribution amplitude
and obeys to the following normalization condition,
\beq
\int_0^1 dx \phi_{B}(x, b=0) &=& \frac{f_{B}}{2 \sqrt{2N_c}}\;.\label{eq:norm}
\eeq
where $b$ is the conjugate space coordinate of transverse momentum $k_T$ and $f_B$
is the decay constant of $B$ meson.
For $B$ meson, the distribution amplitude in the impact $b$
space has been proposed
\beq
\phi_{B}(x,b)&=& N_Bx^2(1-x)^2
\exp\left[-\frac{1}{2}\left(\frac{xm_B}{\omega_b}\right)^2
-\frac{\omega_b^2 b^2}{2}\right] \;,
\eeq
in Refs.~\cite{Keum01:kpi,Lu01:pipi}, where the normalization factor $N_{B}$
is related to the decay constant $f_{B}$ through Eq.~(\ref{eq:norm}).
The shape parameter $\omega_b$ has been fixed at $\omega_b=0.40\pm 0.04$~GeV by using the rich experimental
data on the $B_{u/d}$ mesons with $f_{B_{u/d}}= 0.19$~GeV based on lots of calculations of form factors~\cite{Lu03:form}
and other well-known decay modes of $B_{u/d}$ mesons~\cite{Keum01:kpi,Lu01:pipi}
in the pQCD approach in recent years. By considering the small SU(3) flavor symmetry breaking
effect, the shape parameter $\omega_b$ for $B_s$ meson is taken as $\omega_{B_s}=0.50\pm 0.05$ GeV~\cite{Lu07:bs2mm}.

The light-cone wave function of the light vector meson $K^*$ has been given
in the QCD sum rule method up to twist-3 as~\cite{Ball98:kst}
 \beq
\Phi^{L}_{K^*,\alpha\beta,ij}&\equiv& \langle K^*(P, \epsilon_L)|\bar q(z)_{\beta j} q(0)_{\alpha i} |0\rangle \non
 &=&  \frac{ \delta_{ij}}{\sqrt{2 N_c}} \int^1_0dxe^{ix P\cdot
 z}  \biggl\{ m_{K^*}\, {\epsl}_L \,\phi_{K^*}(x)  +
 {\epsl}_L \, \psl\,\phi_{K^*}^t(x)  + m_{K^*}\, \phi_{K^*}^s(x) \biggr\}_{\alpha\beta}\;,
 \eeq
for longitudinal polarization,
where $\epsilon_{L}$ denotes the longitudinal polarization
vector of $K^*$, satisfying $P \cdot \epsilon_L=0$,
$x$ denotes the momentum
fraction carried by quark in the meson.

The twist-2 distribution amplitude $\phi_{K^*}$ can be parameterized as:
\beq
\phi_{K^*}(x)&=&\frac{3f_{K^*}}{\sqrt{2N_c}} x
(1-x)\left[1+3a_{1K^*}^{||}\, (2x-1)+ a_{2K^*}^{||}\, \frac{3}{2} ( 5(2x-1)^2  - 1 )\right]\;.\label{eq:ldav}
\eeq
And the asymptotic forms of the twist-3 distribution amplitudes
$\phi^{t}_{K^*}$ and $\phi_{K^*}^{s}$ are adopted~\cite{Li05:kphi}:
\beq
\phi^t_{K^*}(x) &=& \frac{3f^T_{K^*}}{2\sqrt {2N_c}}(2x-1)^2,\qquad
\phi^s_{K^*}(x)=-\frac{3f_{K^*}^T}{2\sqrt {2N_c}} (2x-1)~.
\eeq
Here $f_{K^*}$ and $f_{K^*}^T$ are the decay constants of the $K^*$ meson
with longitudinal and transverse polarization, respectively, whose values are
\beq
 f_{K^*} &=& 0.217 \pm 0.005 ~~~~~{\rm GeV}\;,  \qquad f_{K^*}^T = 0.185 \pm 0.010 ~~~~~~{\rm GeV}\;.
\eeq
The Gegenbauer moments are taken from the recent updates~\cite{Ball05:DAs}:
\beq
a_{1K^*}^{||}&=&0.03\pm 0.02,\;\;\;\;\;\;\;\;a_{2K^*}^{||}=0.11 \pm 0.09,\;\;
\eeq

The light-cone wave function of the light scalar $\ks$ has been analyzed
in the QCD sum rule method ~\cite{Cheng06:B2SP}
 \beq
\Phi_{\ks,\alpha\beta,ij}&\equiv& \langle \ks(P)|\bar q(z)_{\beta j} q(0)_{\alpha i} |0\rangle \non
 &=&  \frac{i \delta_{ij}}{\sqrt{2 N_c}} \int^1_0dxe^{ix P\cdot
 z}  \biggl\{\psl\, \phi_{\ks}(x)  +
 m_{\ks}\, \phi_{\ks}^S(x)  + m_{\ks} (\nsl \vsl - 1)\, \phi_{\ks}^T(x) \biggr\}_{\alpha\beta}\;,
 \eeq
where
$n=(1,0,{\bf 0_T})$ and $v=(0,1,{\bf 0_T})$ are the
unit vectors pointing to the plus and minus directions on the light-cone, respectively,
and $x$ denotes the momentum fraction carried by the quark in the $\ks$ meson.

For the light scalar meson $\ks$, its leading twist (twist-2) light-cone distribution amplitude
$\phi_{\ks}(x,\mu)$ can be generally expanded as the Gegenbauer polynomials~\cite{Cheng06:B2SP,Li09:B2Sfm}:
\beq
\phi_{\ks}(x,\mu)&=&\frac{3}{\sqrt{2N_c}}x(1-x)\biggl\{f_{\ks}(\mu)+\bar
f_{\ks}(\mu)\sum_{m=1}^\infty B_m(\mu)C^{3/2}_m(2x-1)\biggr\}, \label{eq:a0-twist-2}
\eeq
where $f_{\ks}(\mu)$ and $\bar f_{\ks}(\mu)$, $B_m(\mu)$, and
$C_m^{3/2}(t)$ are the vector and scalar decay constants,
Gegenbauer moments, and Gegenbauer polynomials, respectively.
There is a relation between the vector and scalar decay constants,
\beq
 \bar f_{\ks} &=& \mu_{\ks} f_{\ks} \;\;\;\;  {\rm and} \;\;\;\;
 \mu_{\ks} = \frac{m_{\ks}}{m_2(\mu)-m_1(\mu)}\;, \label{eq:decay-const}
\eeq
where $m_1$ and $m_2$ are the running current quark masses in the scalar $\ks$.
According to Eq.~(\ref{eq:decay-const}), one can clearly find that the vector decay constant
$f_{\ks}$ is proportional to the mass difference between the constituent
$s$ and $u(d)$ quarks, which will result in $f_{\ks}$ being of order $m_s -m_{u(d)}$. Therefore,
contrary to the case of pseudoscalar mesons,
the contribution from the factorizable diagrams with the emission of $\ks$ will be
largely suppressed.

The values for scalar decay constants and Gegenbauer moments in the distribution amplitudes of $\ks$ have been
estimated at scale $\mu=1~ \mbox{GeV}$ in the scenarios S1 and S2~\cite{Cheng06:B2SP}:
\beq
{\rm S1}: && \quad \bar f_{K_0^*}= -0.300 \pm 0.030~{\rm GeV}, \quad
B_1= 0.58 \pm 0.07, \quad  B_3= -1.20 \pm 0.08,  \non
{\rm S2}: && \quad \bar f_{K_0^*}= 0.445 \pm 0.050~{\rm GeV},
\quad B_1= -0.57 \pm 0.13, \quad  B_3= -0.42 \pm 0.22.
\eeq

As for the twist-3 distribution amplitudes $\phi_{K_0^*}^S$ and
$\phi_{K_0^*}^T$, we here adopt the asymptotic forms
in our numerical calculations as in Ref.~\cite{Cheng06:B2SP}:
\beq
\phi^S_{K_0^*}&=& \frac{1}{2\sqrt {2N_c}}\bar f_{K_0^*},\quad
\phi_{K_0^*}^T=
\frac{1}{2\sqrt {2N_c}}\bar f_{K_0^*}(1-2x).
\eeq
Here, we should stress that the $k_T$ dependence of the distribution amplitudes in the final states
has been neglected, since its contribution is very small as indicated in Refs.~\cite{Li95-96:pQCD}.
The underlying reason is that the contribution
from $k_T$ correlated with a soft dynamics is strongly suppressed by the Sudakov
effect through resummation for the wave function, which is dominated by a collinear dynamics.
Another reason is just that, unfortunately, up to now, the distribution amplitudes with intrinsic $k_T$-dependence
for the above mentioned light mesons $K^*$ and $\ks$ are not available.

\subsection{Perturbative Calculations} \label{ssec:pcalc}

\begin{figure}[!!htb]
\centering
\begin{tabular}{l}
\includegraphics[width=0.9\textwidth]{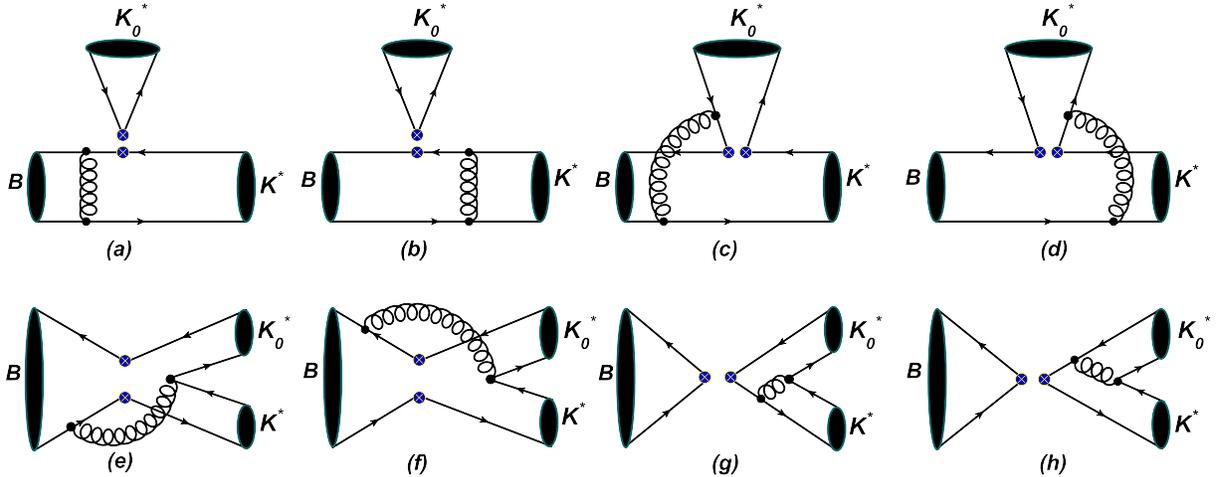}
\end{tabular}
\caption{(Color online) Typical Feynman diagrams contributing to  $B
\to \ks K^*$ decays at leading order. In this figure, $B$ stands for $B_u$,
$B_d$, and $B_s$, respectively. When we exchanged the position of $\ks$ and
 $K^*$, the other eight diagrams
contribute to the considered decay modes will be easily obtained.}
  \label{fig:fig1}
\end{figure}

For the considered $B \to \ks K^*$ decays,
the related weak effective
Hamiltonian $H_{{\rm eff}}$~\cite{Buras96:weak} can be written as
\beq
H_{\rm eff}\, &=&\, {G_F\over\sqrt{2}}
\biggl\{ V^*_{ub}V_{uq} [ C_1(\mu)O_1^{u}(\mu)
+C_2(\mu)O_2^{u}(\mu) ]
 - V^*_{tb}V_{tq} [ \sum_{i=3}^{10}C_i(\mu)O_i(\mu) ] \biggr\}+ {\rm H.c.}\;,
\label{eq:heff}
\eeq
with $q= d$ or $s$, the Fermi constant $G_F=1.16639\times 10^{-5}{\rm
GeV}^{-2}$, Cabibbo-Kobayashi-Maskawa(CKM) matrix elements $V$,
and Wilson coefficients $C_i(\mu)$ at the renormalization scale
$\mu$. The local four-quark
operators $O_i(i=1,\cdots,10)$ are written as
\begin{enumerate}
\item[]{(1) current-current(tree) operators}
\begin{eqnarray}
{\renewcommand\arraystretch{1.5}
\begin{array}{ll}
\displaystyle
O_1^{u}\, =\,
(\bar{q}_\alpha u_\beta)_{V-A}(\bar{u}_\beta b_\alpha)_{V-A}\;,
& \displaystyle
O_2^{u}\, =\, (\bar{q}_\alpha u_\alpha)_{V-A}(\bar{u}_\beta b_\beta)_{V-A}\;;
\end{array}}
\label{eq:operators-1}
\end{eqnarray}

\item[]{(2) QCD penguin operators}
\begin{eqnarray}
{\renewcommand\arraystretch{1.5}
\begin{array}{ll}
\displaystyle
O_3\, =\, (\bar{q}_\alpha b_\alpha)_{V-A}\sum_{q'}(\bar{q}'_\beta q'_\beta)_{V-A}\;,
& \displaystyle
O_4\, =\, (\bar{q}_\alpha b_\beta)_{V-A}\sum_{q'}(\bar{q}'_\beta q'_\alpha)_{V-A}\;,
\\
\displaystyle
O_5\, =\, (\bar{q}_\alpha b_\alpha)_{V-A}\sum_{q'}(\bar{q}'_\beta q'_\beta)_{V+A}\;,
& \displaystyle
O_6\, =\, (\bar{q}_\alpha b_\beta)_{V-A}\sum_{q'}(\bar{q}'_\beta q'_\alpha)_{V+A}\;;
\end{array}}
\label{eq:operators-2}
\end{eqnarray}

\item[]{(3) electroweak penguin operators}
\begin{eqnarray}
{\renewcommand\arraystretch{1.5}
\begin{array}{ll}
\displaystyle
O_7\, =\,
\frac{3}{2}(\bar{q}_\alpha b_\alpha)_{V-A}\sum_{q'}e_{q'}(\bar{q}'_\beta q'_\beta)_{V+A}\;,
& \displaystyle
O_8\, =\,
\frac{3}{2}(\bar{q}_\alpha b_\beta)_{V-A}\sum_{q'}e_{q'}(\bar{q}'_\beta q'_\alpha)_{V+A}\;,
\\
\displaystyle
O_9\, =\,
\frac{3}{2}(\bar{q}_\alpha b_\alpha)_{V-A}\sum_{q'}e_{q'}(\bar{q}'_\beta q'_\beta)_{V-A}\;,
& \displaystyle
O_{10}\, =\,
\frac{3}{2}(\bar{q}_\alpha b_\beta)_{V-A}\sum_{q'}e_{q'}(\bar{q}'_\beta q'_\alpha)_{V-A}\;.
\end{array}}
\label{eq:operators-3}
\end{eqnarray}
\end{enumerate}
with the color indices $\alpha, \ \beta$ and the notations
$(\bar{q}'q')_{V\pm A} = \bar q' \gamma_\mu (1\pm \gamma_5)q'$.
The index $q'$ in the summation of the above operators runs
through $u,\;d,\;s$, $c$, and $b$.
The standard combinations $a_i$ of Wilson coefficients are defined as follows,
  \beq
a_1&=& C_2 + \frac{C_1}{3}\;, \qquad  a_2 = C_1 + \frac{C_2}{3}\;,\quad
 a_i = C_i + \frac{C_{i \pm 1}}{3}(i=3 - 10) \;.
  \eeq
where the upper(lower) sign applies, when $i$ is odd(even).

Similar to $B \to \ks K$ decays~\cite{Liu:scalar}, there are eight types of diagrams
contributing to $B \to \ks K^*$ modes at leading order, as illustrated in Fig.~\ref{fig:fig1}.
They involve two classes of topologies with spectator and annihilation, respectively. Each kind of
topology is classified into factorizable diagrams, in which hard gluon connects the quarks in the
same meson, e.g., Fig.~\ref{fig:fig1} (a) and~\ref{fig:fig1} (b), and nonfactorizable diagrams,
in which hard gluon attaches the quarks in two different mesons, e.g., Fig.~\ref{fig:fig1} (c) and~\ref{fig:fig1} (d).
By calculating these Feynman diagrams, one can get the decay amplitudes of $B \to \ks K^*$ decays.
Because the formulas of $B \to \ks K^*$ are similar to those of $B \to K_0^* K_0^*$~\cite{Liu:scalar},
one can therefore obtain the expressions for all the diagrams just by replacing the corresponding wave functions
and input parameters from $B \to K_0^* K_0^*$. So we do not present the detailed formulas in this paper.

By combining various of contributions from the relevant Feynman diagrams together, the total decay amplitudes
for the considered $B \to \ks K^*$ decays can then read as,
\begin{enumerate}
\item {The total decay amplitudes for charged $B_u$ decays:}
\beq
{\cal A}(B_u \to K^{*+} {\ksb}^0) &=& \lambda_u \biggl[ 
M_{nfa} C_1\biggr]
- \lambda_t \biggl[F_{fs} (a_4- \frac{1}{2} a_{10})
+ F_{fs}^{P2} (a_6 -\frac{1}{2} a_8)  \non
&&  + M_{nfs} (C_3- \frac{1}{2} C_9) + M_{nfs} (C_5 -\frac{1}{2} C_7)
+ M_{nfa}  \non &&
\times  (C_3 + C_9)+ M_{nfa}^{P1} (C_5 + C_7) 
+ f_B F_{fa}^{P2} (a_6 +a_8)\biggr]\;, \label{eq:tda-b2kspk0sb}
\eeq
where $\lambda_{u}= V_{ub}^* V_{ud}$ and $\lambda_t = V_{tb}^* V_{td}$. The
decay amplitude of $B_u \to {\ks}^+ \bar{K}^{*0}$ can be obtained directly from
Eq.~(\ref{eq:tda-b2kspk0sb}) with the replacement of $K^{*} \leftrightarrow {\ks}$, but without the
contributions from the term $F_{fs}^{P2}$. The reason is that the emitted vector $K^*$ meson can not
be produced via the scalar or pseudoscalar current.

\item {The total decay amplitudes for neutral $B_d$ decays:}
\beq
{\cal A}(B_d \to K^{*+} {\ks}^-) &=& \lambda_u \biggl[ M_{nfa} C_2\biggr]-\lambda_t \biggl[
 M_{nfa} (C_4 + C_{10}) + M_{nfa}^{P2} (C_6 + C_8) \non &&
 + M_{nfa}[K^{*+} \leftrightarrow {\ks}^-]
 (C_4 - \frac{1}{2} C_{10})
\non &&
 + M_{nfa}^{P2}[K^{*+} \leftrightarrow {\ks}^-] (C_6 - \frac{1}{2} C_8) \biggr]\;,\label{eq:tda-b2kspk0sm}
\eeq
\beq
{\cal A}(B_d \to K^{*0} {\ksb}^0) &=& -\lambda_t \biggl[ F_{fs}
(a_4 - \frac{1}{2} a_{10}) + F_{fs}^{P2} (a_6 - \frac{1}{2} a_8)
+ (M_{nfs} + M_{nfa} ) \non &&
\times (C_3 -\frac{1}{2} C_9)+(M_{nfs}^{P1}  + M_{nfa}^{P1})(C_5 - \frac{1}{2} C_7)
+ (M_{nfa} \non &&
 + [K^{*0} \leftrightarrow {\ksb}^0] ) (C_4 - \frac{1}{2} C_{10})
 + (M_{nfa}^{P2} + [K^{*0} \leftrightarrow {\ksb}^0] ) \non &&
\times (C_6 - \frac{1}{2} C_8)
 + f_B F_{fa}^{P2} (a_6 - \frac{1}{2}a_8) \biggr]\;.\label{eq:tda-b2ks0k0sb}
\eeq
Similarly, the decay amplitudes of $B_d \to {\ks}^+ K^{*-}$ and $B_d \to {\ks}^0 \bar{K}^{*0}$
can also be obtained easily from Eqs.~(\ref{eq:tda-b2kspk0sm}) and (\ref{eq:tda-b2ks0k0sb}) with
the replacements $K^* \leftrightarrow \ks$, respectively, and with the dropping of term $F_{sf}^{P2}$
for the latter mode.

\item {The total decay amplitudes for $B_s$ decays:}
\beq
{\cal A}(B_s \to {\ks}^+ K^{*-}) &=&
\lambda'_u \biggl[ F_{fs} a_1 + M_{nfs} C_1 + M_{nfa} C_2\biggr]
-\lambda'_t \biggl[ F_{fs} (a_4 + a_{10})  \non &&
+ F_{fs}^{P2} (a_6 +a_8)
+ M_{nfs} (C_3 + C_9) + M_{nfs}^{P1} (C_5 + C_7)
  \non &&
 +  M_{nfa}(C_3 - \frac{1}{2}C_9 + C_4 - \frac{1}{2} C_{10})
+ M_{nfa}[{\ks}^+ \leftrightarrow K^{*-}] \non &&
\times (C_4 + C_{10})
+ M_{nfa}^{P1} (C_5 - \frac{1}{2} C_7)
+ M_{nfa}^{P2} (C_6 - \frac{1}{2} C_8)\non &&
+ M_{nfa}^{P2}[{\ks}^+ \leftrightarrow K^{*-}] (C_6 + C_8)
+ f_{B_s} F_{fa}^{P2} (a_6 - \frac{1}{2} a_8) \biggr]\;,\label{eq:tda-bs2k0spksm}
\eeq
where $\lambda'_{u}= V_{ub}^* V_{us}$ and $\lambda'_t = V_{tb}^* V_{ts}$, and
\beq
{\cal A}(B_s \to {\ks}^0 \bar{K}^{*0}) &=& -\lambda'_t \biggl[ F_{fs}
(a_4 - \frac{1}{2} a_{10}) + F_{fs}^{P2} (a_6 - \frac{1}{2} a_8)
+ (M_{nfs} + M_{nfa} )\non &&
\times (C_3 -\frac{1}{2} C_9)
+(M_{nfs}^{P1}  + M_{nfa}^{P1})(C_5 - \frac{1}{2} C_7)
 + (M_{nfa} \non &&
 + [{\ks}^0 \leftrightarrow \bar{K}^{*0}] )
 (C_4 - \frac{1}{2} C_{10})
 + (M_{nfa}^{P2} + [{\ks}^0 \leftrightarrow \bar{K}^{*0}] )\non &&
 \times (C_6 - \frac{1}{2} C_8)
 +f_{B_s} F_{fa}^{P2} (a_6 -\frac{1}{2} a_8) \biggr]\;.\label{eq:tda-bs2k0s0ks0b}
\eeq
\end{enumerate}
There are other two $B_s$ decay channels, i.e., $B_s \to K^{*+} {\ks}^-$ and $B_s \to K^{*0} {\ksb}^0$, whose decay amplitudes
can be derived from Eqs.~(\ref{eq:tda-bs2k0spksm})
and (\ref{eq:tda-bs2k0s0ks0b}) by the exchange of $\ks \leftrightarrow K^*$, respectively.
Certainly, the $F_{fs}^{P2}$ term has no contribution to them either.
Note that, based on the discussions of the factorizable annihilation contributions $F_{fa}$
in Ref.~\cite{Liu:scalar}, we here neglected this term in the above decay amplitudes
for the considered $B \to \ks K^*$ decays analytically.

\section{Numerical Results and Discussions} \label{sec:randd}

In this section, we will present the theoretical predictions for the branching ratios and CP-violating asymmetries
for those considered $B \to \ks K^*$ decay modes in the pQCD approach.
In numerical calculations, central values of the input parameters will be
used implicitly unless otherwise stated. The relevant QCD scale~({\rm GeV}), masses~({\rm GeV}),
and $B$ meson lifetime({\rm ps}) are the following ~\cite{Keum01:kpi,Lu01:pipi,Beringer12:pdg}
\beq
 \Lambda_{\overline{\mathrm{MS}}}^{(f=4)} &=& 0.250\; , \quad m_W = 80.41\;,
 \quad  m_{B}= 5.28\;,  \quad  m_{B_s}= 5.37\;,\quad m_b = 4.8 \;; \non
  \tau_{B_u} &=& 1.641\;,  \quad \tau_{B_d}= 1.519\;,
   \quad  \tau_{B_s}= 1.497\;,
\quad m_{K^*}= 0.892\;,
\quad m_{K_0^*(1430)}= 1.425\;.
\label{eq:mass}
\eeq
For the CKM matrix elements, we adopt the Wolfenstein
parametrization and the updated parameters $A=0.811$,
 $\lambda=0.22535$, $\bar{\rho}=0.131^{+0.026}_{-0.013}$, and $\bar{\eta}=0.345^{+0.013}_{-0.014}$~\cite{Beringer12:pdg}.

\subsection{Branching Ratios}\label{ssec:cp-brs}

In this subsection, we will analyze the branching ratios of the considered $B \to K_0^* K^*$ decays in
the pQCD approach. For $B \to K_0^* K^*$ decays, the decay rate can be written as
\beq
\Gamma =\frac{G_{F}^{2}m^{3}_{B}}{32 \pi  } (1-2 r_{K_0^*}^2) |{\cal A}(B
\to K_0^* K^*)|^2\;,\label{eq:bqdr}
\eeq
where the corresponding decay amplitudes ${\cal A}$ have been
given explicitly in Eqs.~(\ref{eq:tda-b2kspk0sb}-\ref{eq:tda-bs2k0s0ks0b}).
Using the decay amplitudes obtained in last section, it is straightforward to calculate the
branching ratios with uncertainties as displayed in Table \ref{tab:br-bu} - \ref{tab:br-bs}
for the considered decay modes.
The major errors are induced by the uncertainties of the shape parameters $\omega_b = 0.40 \pm 0.04$~GeV for $B_{u,d}$ decays,
$\omega_{B_s}=0.50 \pm 0.05$~GeV for $B_s$ decays, the scalar decay constant $\bar f_{K_0^*}$ of $\ks$ meson, the decay constants
$f_{K^*}^{(T)}$ of vector $K^*$ meson, the Gegenbauer moments $B_i(i=1, 3)$ for the scalar $K_0^*$,
the Gegenbauer moments $a_{i}(i=1, 2)$ for the vector $K^*$ meson, and CKM matrix elements
$V_i$ ($\bar \rho, \bar \eta$), respectively.

\begin{table*}[htb]
\caption{ The pQCD predictions for the branching ratios of $B_u \to {\ks}^+ \ov{K}^{*0}$ and $K^{*+} {\ksb}^0$ decays
in different scenarios: the first (second)  entry corresponds to S1(S2).}
\label{tab:br-bu}
\begin{center}\vspace{-0.2cm}
{ \begin{tabular}[t]{l|c}
 \hline \hline
 Decay modes     &    Branching ratios                \\
\hline
$B_u \to {\ks}^+ \ov{K}^{*0}$              &$\begin{array}{l}
2.1^{+1.5}_{-0.9}(\omega_{b})
^{+0.4}_{-0.4}(\bar f_{K_0^*})
^{+0.3}_{-0.3}(f^{(T)}_{K^*})
^{+0.1}_{-0.2}(B_{i})
^{+0.2}_{-0.1}(a_{i})
^{+0.1}_{-0.2}(V_i)
^{+0.4}_{-0.2}(a_t)
\times  10^{-7} \\
1.3^{+0.5}_{-0.3}(\omega_{b})
^{+0.3}_{-0.3}(\bar f_{K_0^*})
^{+0.1}_{-0.1}(f^{(T)}_{K^*})
^{+0.1}_{-0.1}(B_{i})
^{+0.1}_{-0.1}(a_{i})
^{+0.0}_{-0.1}(V_i)
^{+0.4}_{-0.2}(a_t)
\times  10^{-6} \end{array} $
\\
\hline 
$B_u \to K^{*+} {\ksb}^0$             &$\begin{array}{l}
6.0^{+1.4}_{-1.0}(\omega_{b})
^{+1.3}_{-1.2}(\bar f_{K_0^*})
^{+0.3}_{-0.3}(f^{(T)}_{K^*})
^{+0.4}_{-0.3}(B_{i})
^{+0.1}_{-0.1}(a_{i})
^{+0.5}_{-0.5}(V_i)
^{+0.8}_{-0.5}(a_t)
\times  10^{-7} \\
1.5^{+0.5}_{-0.3}(\omega_{b})
^{+0.4}_{-0.3}(\bar f_{K_0^*})
^{+0.1}_{-0.1}(f^{(T)}_{K^*})
^{+0.3}_{-0.1}(B_{i})
^{+0.2}_{-0.1}(a_{i})
^{+0.1}_{-0.1}(V_i)
^{+0.5}_{-0.3}(a_t)
\times  10^{-6}\end{array} $
\\
\hline \hline
\end{tabular}}
\end{center}
\end{table*}
\begin{table*}[htb]
\caption{ Same as Table~\ref{tab:br-bu} but for neutral $B_d \to \ks K^*$ decays in both scenarios.}
\label{tab:br-bd}
\begin{center}\vspace{-0.2cm}
{ \begin{tabular}[t]{l|c}
 \hline \hline
 Decay modes     &    Branching ratios                \\
\hline
$B_d \to {\ks}^0 \ov{K}^{*0} + K^{*0} {\ksb}^0$              &$\begin{array}{l}
6.4^{+0.5}_{-0.6}(\omega_{b})
^{+1.3}_{-1.2}(\bar f_{K_0^*})
^{+0.7}_{-0.7}(f^{(T)}_{K^*})
^{+1.6}_{-1.5}(B_{i})
^{+0.8}_{-0.8}(a_{i})
^{+0.2}_{-0.3}(V_i)
^{+0.2}_{-0.5}(a_t)
\times  10^{-7} \\
5.9^{+0.7}_{-0.6}(\omega_{b})
^{+1.4}_{-1.2}(\bar f_{K_0^*})
^{+0.6}_{-0.4}(f^{(T)}_{K^*})
^{+5.9}_{-3.2}(B_{i})
^{+0.5}_{-0.4}(a_{i})
^{+0.2}_{-0.3}(V_i)
^{+0.7}_{-0.6}(a_t)
\times  10^{-7}
\end{array} $
\\
\hline
$B_d/\bar B_d \to {\ks}^0 \ov{K}^{*0} (K^{*0}  {\ksb}^{0})$              &$\begin{array}{l}
5.0^{+2.2}_{-1.3}(\omega_{b})
^{+1.1}_{-1.0}(\bar f_{K_0^*})
^{+0.6}_{-0.6}(f^{(T)}_{K^*})
^{+0.4}_{-0.3}(B_{i})
^{+0.2}_{-0.1}(a_{i})
^{+0.1}_{-0.3}(V_i)
^{+0.5}_{-0.5}(a_t)
\times  10^{-7} \\
2.3^{+0.8}_{-0.5}(\omega_{b})
^{+0.5}_{-0.5}(\bar f_{K_0^*})
^{+0.1}_{-0.1}(f^{(T)}_{K^*})
^{+0.4}_{-0.2}(B_{i})
^{+0.2}_{-0.2}(a_{i})
^{+0.0}_{-0.1}(V_i)
^{+0.7}_{-0.5}(a_t)
\times  10^{-6}
 \end{array} $
\\
\hline \hline
 Decay modes     &    Branching ratios ($10^{-6}$)               \\
\hline
$B_d \to {\ks}^+ K^{*-} + K^{*+} {\ks}^-$             &$\begin{array}{l}
2.8^{+0.2}_{-0.3}(\omega_{b})
^{+0.6}_{-0.6}(\bar f_{K_0^*})
^{+0.1}_{-0.2}(f^{(T)}_{K^*})
^{+0.4}_{-0.4}(B_{i})
^{+0.1}_{-0.0}(a_{i})
^{+0.1}_{-0.2}(V_i)
^{+0.0}_{-0.1}(a_t) \\
1.1^{+0.1}_{-0.1}(\omega_{b})
^{+0.2}_{-0.3}(\bar f_{K_0^*})
^{+0.0}_{-0.1}(f^{(T)}_{K^*})
^{+1.1}_{-0.6}(B_{i})
^{+0.1}_{-0.1}(a_{i})
^{+0.0}_{-0.1}(V_i)
^{+0.0}_{-0.1}(a_t) \end{array} $
\\
\hline
$B_d/\bar B_d \to {\ks}^+ K^{*-}$              &$\begin{array}{l}
4.5^{+0.4}_{-0.4}(\omega_{b})
^{+0.9}_{-0.9}(\bar f_{K_0^*})
^{+0.2}_{-0.2}(f^{(T)}_{K^*})
^{+0.6}_{-0.6}(B_{i})
^{+0.3}_{-0.3}(a_{i})
^{+0.2}_{-0.3}(V_i)
^{+0.1}_{-0.1}(a_t)  \\
2.9^{+0.3}_{-0.4}(\omega_{b})
^{+0.6}_{-0.7}(\bar f_{K_0^*})
^{+0.2}_{-0.1}(f^{(T)}_{K^*})
^{+2.0}_{-1.4}(B_{i})
^{+0.1}_{-0.1}(a_{i})
^{+0.2}_{-0.2}(V_i)
^{+0.1}_{-0.2}(a_t) \end{array} $
\\
\hline
$B_d/\bar B_d \to K^{*+} {\ks}^{-}$              &$\begin{array}{l}
1.7^{+0.2}_{-0.2}(\omega_{b})
^{+0.3}_{-0.3}(\bar f_{K_0^*})
^{+0.1}_{-0.1}(f^{(T)}_{K^*})
^{+0.2}_{-0.2}(B_{i})
^{+0.1}_{-0.1}(a_{i})
^{+0.1}_{-0.2}(V_i)
^{+0.1}_{-0.1}(a_t)  \\
1.1^{+0.1}_{-0.2}(\omega_{b})
^{+0.2}_{-0.3}(\bar f_{K_0^*})
^{+0.1}_{-0.0}(f^{(T)}_{K^*})
^{+0.6}_{-0.5}(B_{i})
^{+0.0}_{-0.1}(a_{i})
^{+0.1}_{-0.1}(V_i)
^{+0.0}_{-0.1}(a_t) \end{array} $
\\
\hline \hline
\end{tabular}}
\end{center}
\end{table*}
\begin{table*}[htb]
\caption{ Same as Table~\ref{tab:br-bu} but for strange $B_s \to \ks K^*$ decays in both scenarios.}
\label{tab:br-bs}
\begin{center}\vspace{-0.2cm}
{ \begin{tabular}[t]{l|c}
 \hline \hline
 Decay modes     &    Branching ratios ($10^{-5}$)               \\
\hline
$B_s \to {\ks}^0 \ov{K}^{*0} + K^{*0} {\ksb}^0$              &$\begin{array}{l}
1.3^{+0.2}_{-0.1}(\omega_{bs})
^{+0.3}_{-0.2}(\bar f_{K_0^*})
^{+0.2}_{-0.1}(f^{(T)}_{K^*})
^{+0.4}_{-0.3}(B_{i})
^{+0.2}_{-0.1}(a_{i})
^{+0.0}_{-0.0}(V_i)
^{+0.1}_{-0.1}(a_t)  \\
1.3^{+0.2}_{-0.2}(\omega_{bs})
^{+0.3}_{-0.3}(\bar f_{K_0^*})
^{+0.1}_{-0.1}(f^{(T)}_{K^*})
^{+1.2}_{-0.7}(B_{i})
^{+0.1}_{-0.1}(a_{i})
^{+0.0}_{-0.0}(V_i)
^{+0.1}_{-0.2}(a_t) \end{array} $
\\
\hline
$B_s/\bar B_s \to {\ks}^0 \ov{K}^{*0} (K^{*0} {\ksb}^{0})$              &$\begin{array}{l}
0.9^{+0.3}_{-0.2}(\omega_{bs})
^{+0.2}_{-0.2}(\bar f_{K_0^*})
^{+0.1}_{-0.1}(f^{(T)}_{K^*})
^{+0.1}_{-0.1}(B_{i})
^{+0.1}_{-0.1}(a_{i})
^{+0.0}_{-0.0}(V_i)
^{+0.1}_{-0.1}(a_t)  \\
5.4^{+1.5}_{-0.9}(\omega_{bs})
^{+1.3}_{-1.1}(\bar f_{K_0^*})
^{+0.2}_{-0.2}(f^{(T)}_{K^*})
^{+0.9}_{-0.5}(B_{i})
^{+0.9}_{-0.6}(a_{i})
^{+0.0}_{-0.0}(V_i)
^{+1.7}_{-1.1}(a_t) \end{array} $
\\
\hline \hline
 Decay modes     &    Branching ratios ($10^{-5}$)               \\
\hline
$B_s \to {\ks}^+ K^{*-} + K^{*+} {\ks}^-$             &$\begin{array}{l}
1.3^{+0.2}_{-0.1}(\omega_{bs})
^{+0.3}_{-0.2}(\bar f_{K_0^*})
^{+0.1}_{-0.1}(f^{(T)}_{K^*})
^{+0.4}_{-0.2}(B_{i})
^{+0.1}_{-0.1}(a_{i})
^{+0.1}_{-0.0}(V_i)
^{+0.1}_{-0.1}(a_t)  \\
1.5^{+0.4}_{-0.3}(\omega_{bs})
^{+0.4}_{-0.3}(\bar f_{K_0^*})
^{+0.1}_{-0.0}(f^{(T)}_{K^*})
^{+1.0}_{-0.4}(B_{i})
^{+0.1}_{-0.0}(a_{i})
^{+0.1}_{-0.0}(V_i)
^{+0.3}_{-0.2}(a_t) \end{array} $
\\
\hline
$B_s/\bar B_s \to {\ks}^+ K^{*-}$              &$\begin{array}{l}
0.9^{+0.3}_{-0.1}(\omega_{bs})
^{+0.2}_{-0.2}(\bar f_{K_0^*})
^{+0.1}_{-0.1}(f^{(T)}_{K^*})
^{+0.2}_{-0.1}(B_{i})
^{+0.0}_{-0.0}(a_{i})
^{+0.0}_{-0.0}(V_i)
^{+0.1}_{+0.1}(a_t)  \\
4.0^{+1.3}_{-0.7}(\omega_{bs})
^{+1.0}_{-0.8}(\bar f_{K_0^*})
^{+0.2}_{-0.1}(f^{(T)}_{K^*})
^{+0.8}_{-0.3}(B_{i})
^{+0.6}_{-0.4}(a_{i})
^{+0.1}_{-0.1}(V_i)
^{+1.4}_{-0.8}(a_t)  \end{array} $
\\
\hline
$B_s/\bar B_s \to K^{*+} {\ks}^{-}$              &$\begin{array}{l}
0.9^{+0.2}_{-0.1}(\omega_{bs})
^{+0.2}_{-0.2}(\bar f_{K_0^*})
^{+0.1}_{-0.1}(f^{(T)}_{K^*})
^{+0.1}_{-0.1}(B_{i})
^{+0.2}_{-0.1}(a_{i})
^{+0.0}_{-0.0}(V_i)
^{+0.2}_{-0.1}(a_t)  \\
6.1^{+1.5}_{-0.9}(\omega_{bs})
^{+1.4}_{-1.3}(\bar f_{K_0^*})
^{+0.3}_{-0.2}(f^{(T)}_{K^*})
^{+1.0}_{-0.6}(B_{i})
^{+1.1}_{-0.9}(a_{i})
^{+0.1}_{-0.1}(V_i)
^{+1.9}_{-1.2}(a_t) \end{array} $
\\
\hline \hline
\end{tabular}}
\end{center}
\end{table*}

Based on the above numerical results of the branching ratios given at leading order
in the pQCD approach for the considered decay modes, some remarks are as follows:

\begin{enumerate}
\item[(1)] Generally speaking, the theoretical predictions for the considered decays
in the pQCD approach have relatively large errors arising from the still large uncertainties
of many input parameters. Furthermore, the numerical results for the branching ratios
suffer more from the errors induced by the less constrained hadronic parameters of
the light scalar $\ks$, such as the scalar decay constant $\bar f_{\ks}$ and the Gegenbauer
coefficients $B_i(i=1,3)$. Additionally, in this work, as displayed in the above tables,
the higher order contributions are also simply investigated by exploring the variation
of the hard scale $t_{\rm max}$, i.e., from $0.8t$ to $1.2t$ (not changing $1/b_i, i= 1,2,3$),
in the hard kernel, which have been counted into one of the source of theoretical
uncertainties.

\item[(2)]
The pQCD predictions for the CP-averaged $Br(B_u \to {\ks}^+
\overline{K}^{*0})$ and $Br(B_u \to K^{*+} {\ksb}^0)$ are in the order
of $10^{-6}$ in S2, which are lager than those in S1,
and can be tested by the  future $B$ physics experiments.
Moreover, one can define the ratios of the branching ratios
of the same decay mode but in different scenarios as the following,
 \beq
 \frac{Br(B_u \to {\ks}^+ \overline{K}^{*0})_{\rm S2}}{Br(B_u \to {\ks}^+ \overline{K}^{*0})_{\rm S1}}
 &=& 6.2\;, \qquad
 \frac{Br(B_u \to K^{*+} {\ksb}^{0})_{\rm S2}}{Br(B_u \to K^{*+} {\ksb}^{0})_{\rm S1}}
 = 2.5\;;
 \eeq
where the central values are quoted for clarification.
The above two patterns imply the different QCD dynamics involved in the corresponding decay
channels, which can be tested with the future precision measurements.

\item[(3)]
For the neutral $B_d$ decays, which include the pure penguin contribution modes,
i.e., $B_d \to K^{*0} {\ksb}^0$ and $B_d \to {\ks}^0 \overline{K}^{*0}$, and the pure annihilation
contribution channels, i.e., $B_d \to K^{*+} {\ks}^-$ and $B_d \to {\ks}^+ K^{*-}$, respectively.
The analysis for these four decay modes are a little complicated, which is just because both $B_d$
and $\bar B_d$
can decay into the same final states simultaneously, in other words, the final states
in the considered $B_d$ decays are not the CP-eigensates. Due to the $B_d-\bar B_d$ mixing, it is very difficult for us to distinguish the $B_d$
from the $\bar B_d$. However, fortunately, it is easy to identify the final states in the considered
decays. We therefore sum up $B_d/\bar B_d \to K^{*0} {\ksb}^0$ as one channel, and $B_d/\bar B_d \to
{\ks}^0 \overline{K}^{*0}$ as another. Similarly, we will have $B_d/\bar B_d \to K^{*+} {\ks}^-$ as one mode, and $B_d/\bar B_d \to {\ks}^+ K^{*-}$ as another. Moreover, following the convention by the
experimental measurements~\cite{Beringer12:pdg,Amhis12:hfag}, we also define the averaged quantity of the two channels, i.e., $B_d \to K^{*0} {\ksb}^0 + {\ks}^0 \overline{K}^{*0}$
and $B_d \to K^{*+} {\ks}^{-} + {\ks}^+ K^{*-}$.
The same phenomena will also occur in the decays of the $B_s$ meson.

\item[(4)]
The theoretical predictions on the branching ratios of the $B_d$ meson decays in
the pQCD approach have been presented in Table~\ref{tab:br-bd}.
For the pure penguin $B_d/\bar B_d \to {\ks}^0 \overline{K}^{*0}$, $B_d/\bar B_d \to K^{*0} {\ksb}^0$, and
$B_d \to {\ks}^0 \overline{K}^{*0} + K^{*0} {\ksb}^0$ channels,
the pQCD predictions for the former two decays show that the branching ratios (about $ 2 \times 10^{-6}$)
in S2 are larger than that (about $5 \times 10^{-7}$) in S1, which results in the ratio $Br(B_d/\bar B_d \to {\ks}^0
\overline{K}^{*0}(K^{*0} {\ksb}^0))_{\rm S2}/Br(B_d/\bar B_d \to {\ks}^0 \overline{K}^{*0}(K^{*0} {\ksb}^0))_{\rm S1} \approx 4.6$;
while the pQCD predictions for the latter one in both scenarios are similar, which leads to the ratio
$Br(B_d \to K^{*0} {\ksb}^0 + {\ks}^0 \overline{K}^{*0})_{\rm S1}/
Br(B_d \to K^{*0} {\ksb}^0 + {\ks}^0 \overline{K}^{*0})_{\rm S2} \approx 1.1$. Note that due to the charge conjugation
between the pure penguin channels $B_d \to {\ks}^0 \overline{K}^{*0}$ and $\bar B_d \to {\ksb}^0 K^{*0}$ and the domination
of the real contributions arising from the factorizable emission diagrams, e.g., Fig.~\ref{fig:fig1} (a) and (b),
in the considered channels,
which give the same branching ratios for $B_d/\bar B_d \to {\ks}^0 \overline{K}^{*0}$
and $B_d/\bar B_d \to {\ksb}^0 K^{*0}$ modes,
as presented in Table~\ref{tab:br-bd}. Certainly, the similar phenomena will also appear in the related
$B_s$ meson decays.

As shown in Eq.~(1), only a preliminary upper limit for $B^0 \to {\kst}^0 \ov{K}^{*0}$ decay
is available now. By comparison, one can easily find from Table~\ref{tab:br-bd}
that the pQCD predictions in both scenarios are all consistent with this upper limit.
The branching ratios in the order of $10^{-6}$ and above are expected to be tested in the near future $B$
meson experiments.

\item[(5)]
For the pure annihilation $B_d/\bar B_d \to {\ks}^+ K^{*-}$, $B_d/\bar B_d \to K^{*+} {\ks}^-$,
and $B_d \to {\ks}^+ K^{*-} + K^{*+} {\ks}^-$ channels, as listed in Table~\ref{tab:br-bd}, the pQCD predictions for the
branching ratios in both scenarios are in order of $10^{-6}$.
Furthermore, one can find that the numerical pQCD results
for the branching ratios in S1 are clearly larger than those in S2, which are rather different from the situation
of $B_d \to {\ks}^0 \overline{K}^{*0}$ and $K^{*0} {\ksb}^0$ decays. Although the charge conjugation also exists
in the channels $B_d \to K^{*+} {\ks}^-$ and $\bar B_d \to K^{*-} {\ks}^+$, the interference between tree
and penguin topologies makes the decay amplitudes for the considered modes different from those for
the $B_d$ meson decaying into two neutral final states, which therefore give different branching ratios
for $B_d/\bar B_d \to {\ks}^+ K^{*-}$ and $B_d/\bar B_d \to K^{*+} {\ks}^-$ as exhibited in Table~\ref{tab:br-bd}.

Additionally, one can define the interesting ratios among the same decay modes but in different scenarios,
 \beq
  \frac{Br(B_d \to {\ks}^+ K^{*-} + K^{*+} {\ks}^-)_{\rm S1}}{Br(B_d \to {\ks}^+ K^{*-} + K^{*+} {\ks}^-)_{\rm S2}} &=& 2.5\;,
  \eeq
 \beq
  \frac{Br(B_d/\bar B_d \to {\ks}^+ K^{*-})_{\rm S1}}{Br(B_d/\bar B_d \to {\ks}^+ K^{*-})_{\rm S2}} &=& 1.6\;, \qquad
  \frac{Br(B_d/\bar B_d \to K^{*+} {\ks}^-)_{\rm S1}}{Br(B_d/\bar B_d \to K^{*+} {\ks}^-)_{\rm S2}} = 1.5\;,
 \eeq
where only the central values of the branching ratios are considered for clarification.

Very recently, LHCb~\cite{Powell11:LHCb} and CDF~\cite{Ruffini11:CDF} Collaborations have measured
the pure annihilation modes of charmless hadronic $B$ meson decays, such as $B_d \to K^+ K^-$ and
$B_s \to \pi^+ \pi^-$, respectively. It is therefore believed that such large decay rates (about
$(1 \sim 5) \times 10^{-6}$) for
the considered pure annihilation decays in this paper could be tested by the ongoing LHC experiments
and/or the forthcoming Super-$B$ factory in the near future. If the numerical results of the pure annihilation
decays can be confirmed by the future measurements at the predicted level, on one hand, which will provide
much more evidences to support the successful pQCD approach in calculating the annihilation diagrams;
on the other hand, which will provide more important information on the sizable annihilation contributions
in heavy $B$ meson physics and further shed light on
the underlying mechanism of the annihilated $B$ meson decays.

\item[(6)]
For the considered $B_s$ meson decays, all the predicted branching ratios are in the range
of $(1 \sim 6) \times 10^{-5}$, which can be seen in Table~\ref{tab:br-bs}
and will be tested by the LHC experiments. In terms of the channels
with the neutral final states, which are the pure penguin induced decays, the branching ratios for
the averaged channel $B_s \to {\ks}^0 \overline{K}^{*0} + K^{*0} {\ksb}^0$ are equal to each other
in two scenarios.
The branching ratios for the two summed channels, $B_s/\bar B_s \to {\ks}^0 \overline{K}^{*0}$
and $B_s/\bar B_s \to K^{*0} {\ksb}^0$ in S2, however, are larger than those in S1 with a factor about six.

The pQCD predictions for the branching ratios for the averaged channel $B_s \to {\ks}^+ K^{*-} + K^{*+} {\ks}^-$
are similar in size in both scenarios.
Analogous to the decays with the neutral final states, the pQCD results for
$B_s/\bar B_s \to {\ks}^+ K^{*-}$ and $B_s/\bar B_s \to K^{*+} {\ks}^-$ in S2 are larger than those in S1
with a factor about 4 and 6, respectively.
These results are expected to be examined by the measurements in the future.

\item[(7)]
For the considered pure penguin decays, $B_{d/s} \to {\ks}^0 \overline{K}^{*0}$ and $K^{*0} {\ksb}^0$,
we get the ratios in two scenarios between the branching ratios of $B_d$ and $B_s$ decays in the pQCD approach,
\beq
\frac{\tau_{B_d}}{\tau_{B_s}} \cdot \frac{Br(B_s \to {\ks}^0 \overline{K}^{*0} + K^{*0} {\ksb}^0)}{Br(B_d \to {\ks}^0 \overline{K}^{*0} + K^{*0} {\ksb}^0)}
 &=& 20.6 \;, \qquad
\frac{\tau_{B_d}}{\tau_{B_s}} \cdot \frac{Br(B_s/\bar B_s \to {\ks}^0 \overline{K}^{*0} (K^{*0} {\ksb}^0))}{Br(B_d/ \bar B_d \to {\ks}^0 \overline{K}^{*0} (K^{*0} {\ksb}^0))}
 = 18.3\;;
\eeq
in S1, and
\beq
\frac{\tau_{B_d}}{\tau_{B_s}} \cdot \frac{Br(B_s \to {\ks}^0 \overline{K}^{*0} + K^{*0} {\ksb}^0)}{Br(B_d \to {\ks}^0 \overline{K}^{*0} + K^{*0} {\ksb}^0)}
 &=& 22.4 \;, \qquad
\frac{\tau_{B_d}}{\tau_{B_s}} \cdot \frac{Br(B_s/\bar B_s \to {\ks}^0 \overline{K}^{*0} (K^{*0} {\ksb}^0))}{Br(B_d/ \bar B_d \to {\ks}^0 \overline{K}^{*0} (K^{*0} {\ksb}^0))}
 = 23.8\;;
\eeq
in S2, in which the central values of the branching ratios are quoted.  From the analytical expressions for the decay amplitudes of these $B_d$ and $B_s$ modes, e.g., Eqs.~(\ref{eq:tda-b2ks0k0sb})
and (\ref{eq:tda-bs2k0s0ks0b}), one can easily find that the main difference is just from the involved CKM factors $\lambda_t$ and
$\lambda'_t$ with $|\lambda'_t/\lambda_t|^2 = 22.5$.

\item[(8)]
Frankly speaking, the measurements at the experimental aspect are not yet available
up to now. We therefore can not make any judgements on whether the scenario 1 or scenario 2 of
the scalar $\ks$ is favored by the considered decays.
The pQCD predictions for the branching ratios of the considered $B \to \ks K^*$ decays will be tested by the LHC experiments
and/or forthcoming Super-$B$ facility.

\end{enumerate}

Here, based on the numerical calculations of the branching ratios, we also examine the effects
coming from the annihilation diagrams.
In those considered $B \to \ks K^*$ decays, when the annihilation contributions are not taken into account,
the relevant predictions on the branching ratios in the pQCD approach are as follows:
 \beq
Br(B_u \to {\ks}^+ \ov{K}^{*0}) &=& 7.8 \times 10^{-7}, \qquad
Br(B_u \to K^{*+} {\ksb}^{0}) = 6.1
\times 10^{-7} \;,
\eeq
 \beq
Br(B_d \to {\ks}^0 \ov{K}^{*0} + K^{*0} {\ksb}^0) &=& 1.7 \times 10^{-8}, \quad
Br(B_d/\bar B_d \to {\ks}^0 \ov{K}^{*0}(K^{*0}
{\ksb}^0)) = 1.3 \times 10^{-6}\;,
 \eeq
 \beq
Br(B_s \to {\ks}^0 \ov{K}^{*0} + K^{*0} {\ksb}^0) &=& 3.2 \times 10^{-7}, \quad
Br(B_s/\bar B_s \to {\ks}^0 \ov{K}^{*0}(K^{*0}
{\ksb}^0)) = 2.4 \times 10^{-5}\;,
 \eeq
 \beq
Br(B_s \to {\ks}^+ K^{*-} + K^{*+} {\ks}^-) &=& 0.3 \times 10^{-5},
\eeq
\beq
Br(B_s/\bar B_s \to {\ks}^+ K^{*-}) &=& 2.2 \times 10^{-5}, \quad
Br(B_s/\bar B_s \to K^{*+}
{\ks}^-)=  2.1 \times 10^{-5}\;;
\eeq
in scenario 1, and
 \beq
Br(B_u \to {\ks}^+ \ov{K}^{*0}) &=& 1.8 \times 10^{-6}, \qquad
Br(B_u \to K^{*+} {\ksb}^{0}) = 1.5
\times 10^{-6}\;,
 \eeq
 \beq
Br(B_d \to {\ks}^0 \ov{K}^{*0} + K^{*0} {\ksb}^0) &=& 3.4 \times 10^{-8}, \quad
Br(B_d/\bar B_d \to {\ks}^0 \ov{K}^{*0}(K^{*0}
{\ksb}^0)) = 3.0 \times 10^{-6}\;,
 \eeq
 \beq
Br(B_s \to {\ks}^0 \ov{K}^{*0} + K^{*0} {\ksb}^0) &=& 3.0 \times 10^{-7}, \quad
Br(B_s/\bar B_s \to {\ks}^0 \ov{K}^{*0}(K^{*0}
{\ksb}^0)) = 5.4 \times 10^{-5}\;,
 \eeq
 \beq
Br(B_s \to {\ks}^+ K^{*-} + K^{*+} {\ks}^-) &=& 0.7 \times 10^{-5},
 \eeq
 \beq
Br(B_s/\bar B_s \to {\ks}^+ K^{*-})&=& 5.0 \times 10^{-5}, \quad
Br(B_s/\bar B_s \to K^{*+}
{\ks}^-) =  4.9 \times 10^{-5}\;.
\eeq
in scenario 2, in which only the central values are considered for estimating the contributions
arising from the annihilation diagrams in various decay channels. By comparison, one can easily
find the following points:
\begin{itemize}
\item[(1)]
For the charged $B_u$ decays, the weak annihilation contributions play more important roles in the
$B_u \to {\ks}^+ \overline{K}^{*0}$ than that in the $B_u \to K^{*+} {\ksb}^0$.

\item[(2)]
For the pure penguin $B_d$ and $B_s$ decays, apart from the $B_s/\bar B_s \to {\ks}^0 \overline{K}^{*0}(K^{*0} {\ksb}^0)$
in S2, other channels are basically dominated by the weak annihilation contributions, particularly, for
$B_{d/s} \to {\ks}^0 \overline{K}^{*0} + K^{*0} {\ksb}^0$ modes.

\item[(3)]
For other $B_s$ decays, the significant contributions given by the weak annihilation diagrams can also be clearly
observed. Of course, the reliability
of the contributions from the annihilation diagrams to these considered decays calculated in the pQCD approach
will be examined by the relevant experiments in the future.
\end{itemize}

\subsection{CP-violating Asymmetries} \label{ssec:cpv}

Now we turn to the evaluations of the CP-violating asymmetries of $B
\to K_0^* K^*$ decays in the pQCD approach.
For the charged $B_u$ meson decays, the direct CP violation
$\acp^{\rm dir}$ can be defined as,
 \beq
\acp^{\rm dir} =  \frac{|\overline{\cal A}_f|^2 - |{\cal A}_f|^2}{
 |\overline{\cal A}_f|^2+|{\cal A}_f|^2},
\label{eq:acp1}
\eeq
where ${\cal A}_f$ stands for the decay amplitude of $B_u \to {\ks}^+ \ov{K}^{*0}$
and $B_u \to K^{*+} {\ksb}^0$, respectively, while $\ov{{\cal A}}_f$ denotes the
charge conjugation one correspondingly. Using Eq.~(\ref{eq:acp1}), we find the following pQCD predictions
(in unit of $10^{-2}$):
\beq
\acp^{\rm dir}(B_u \to {\ks}^+ \overline{K}^{*0}) &=& \left\{ \begin{array}{ll}
-32.6^{+10.5}_{-11.3}(\omega_{b})
^{+0.7}_{-0.8}(\bar f_{K_0^*})
^{+3.1}_{-4.0}(f_{K^*})
^{+3.6}_{-4.8}(B_{i})
^{+5.4}_{-1.7}(a_{i})
^{+1.1}_{-2.4}(V_i)
^{+2.1}_{-3.9}(a_t) &
\hspace{0.1cm} ({\rm S1}) \\
-34.9^{+5.0}_{-4.5}(\omega_{b})
^{+0.5}_{-0.4}(\bar f_{K_0^*})
^{+1.6}_{-1.5}(f_{K^*})
^{+6.9}_{-9.0}(B_{i})
^{+1.5}_{-1.6}(a_{i})
^{+1.4}_{-2.2}(V_i)
^{+1.6}_{-0.2}(a_t) &
 \hspace{0.1cm} ({\rm S2}) \\ \end{array} \right.  \label{eq:adir-k0spksb-u}\;, \\
\acp^{\rm dir}(B_u \to {K}^{*+} {\ksb}^0) &=& \left\{ \begin{array}{ll}
\hspace{0.26cm}43.6^{+3.7}_{-2.2}(\omega_{b})
^{+1.1}_{-1.3}(\bar f_{K_0^*})
^{+2.0}_{-1.7}(f_{K^*})
^{+13.0}_{-13.7}(B_{i})
^{+2.8}_{-2.6}(a_{i})
^{+1.7}_{-1.6}(V_i)
^{+2.0}_{-2.7}(a_t) &
\hspace{0.1cm} ({\rm S1}) \\
-67.9^{+4.9}_{-5.2}(\omega_{b})
^{+0.8}_{-1.0}(\bar f_{K_0^*})
^{+1.7}_{-1.6}(f_{K^*})
^{+16.0}_{-14.9}(B_{i})
^{+1.2}_{-0.6}(a_{i})
^{+1.9}_{-3.8}(V_i)
^{+3.3}_{-1.3}(a_t) &
\hspace{0.1cm} ({\rm S2}) \\ \end{array} \right.  \label{eq:adir-kspk0sb-u}\;;
\eeq
Note that these two channels exhibit large direct CP-violating asymmetries in both scenarios
in the pQCD approach, which indicates that the contribution of the penguin diagrams is sizable.
Combining the large CP-averaged branching ratios(${\cal O}(10^{-6})$) in S2 with the large
CP violations, which could be clearly detected in $B$ factories and LHC experiments and will
provide important information on further understanding of the QCD dynamics involved in the
considered scalar $\ks$.

As for the CP-violating asymmetries for the neutral $B_{d/s} \to \ks K^*$ decays, the effects of
$B_{d/s}-\bar{B}_{d/s}$ mixing should be considered. Firstly,
for $B_{d/s}/\bar{B}_{d/s} \to {\ks}^0 \ov{K}^{*0}$ and $K^{*0} {\ksb}^0$ decays,
they will not exhibit CP violation in both scenarios 1 and 2,
since they involve the pure penguin
contributions at the leading order in the SM, which can be seen from the decay
amplitudes as given in Eqs.~(\ref{eq:tda-b2ks0k0sb}) and (\ref{eq:tda-bs2k0s0ks0b}).
If the measurements from experiments for the direct CP asymmetries
$\acp^{\rm dir}$ in $B_{d/s} \to {\ks}^0 \ov{K}^{*0}$ and $K^{*0} {\ksb}^0$ decays exhibit obviously nonzero,
which will indicate the existence of new physics beyond the SM and will provide a
very promising place to look for this exotic effect.

However, the study of CP-violation for $B_{d/s}\to {K_0^*}^+ K^{*-}$
and $K^{*+} {K_0^*}^-$ becomes more complicated as
${\ks}^+ \ov{K}^{*-}$ and $K^{*+} {\ksb}^-$
are not CP eigenstates.
The time-dependent CP
asymmetries for $B_{d/s} \to {\ks}^{\pm} K^{*\mp}$ decays are thus given by
\beq
a_{CP} &\equiv& \frac{\Gamma\left
(\bar{B}_{d/s}(\Delta t) \to {\ks}^{\pm} K^{*\mp}\right) -
\Gamma\left(B_{d/s}(\Delta t) \to {\ks}^{\pm} K^{*\mp}\right )}{ \Gamma\left
(\bar{B}_{d/s}(\Delta t) \to {\ks}^{\pm} K^{*\mp}\right ) + \Gamma\left
(B_{d/s}(\Delta t) \to {\ks}^{\pm} K^{*\mp}\right ) }\non
&=& (\acp^{\rm dir} \pm \Delta \acp^{\rm dir})
\cos(\Delta m_{(d/s)}  \Delta t) + (\acp^{\rm mix} \pm \Delta \acp^{\rm mix})
\sin (\Delta m_{(d/s)} \Delta
t), \label{eq:acp-def} \eeq where $\Delta m_{(d/s)}$ is the mass
difference between the two neutral $B_{d/s}$ mass eigenstates, $\Delta
t =t_{CP}-t_{tag} $ is the time difference between the tagged
$B_{d/s}$ ($\bar{B}_{d/s}$) and the accompanying
$\bar{B}_{d/s}$ ($B_{d/s}$) with opposite $b$ flavor
decaying to the final CP-eigenstate ${\ks}^{\pm} K^{*\mp}$ at the time $t_{CP}$.
The quantities $\acp^{\rm
dir} ({\cal C}_f)$ 
and $\acp^{\rm mix} ({\cal S}_f)$ parameterize flavor-dependent
direct CP violation and mixing-induced CP violation, respectively,
and the parameters $\Delta\acp^{\rm dir}$ and $\Delta\acp^{\rm mix}$ are
related CP-conserving quantities: $\Delta\acp^{\rm dir}$ describes the asymmetry
between the rates $\Gamma(B_{d/s} \to {K_0^*}^{+} K^{*-})+
\Gamma(\bar B_{d/s} \to {K_0^*}^- K^{*+} )$ and $\Gamma(B_{d/s} \to {K_0^*}^{-} K^{*+})+
\Gamma(\bar B_{d/s} \to {K_0^*}^+ K^{*-} )$, while $\Delta\acp^{\rm mix}$ measures the strong
phase difference between the amplitudes contributing to $B_{d/s} \to {\ks}^{\pm} K^{*\mp}$ decays.
Here, we should stress that in the definition
of the above equation, i.e., Eq.~(\ref{eq:acp-def}), the effects arising from
the width difference of $B_s$ meson have been neglected for simplicity.

Following Ref.~\cite{CKMfitter}, we define the transition amplitudes, for $B_d$ decays for example, as follows,
\beq
{\cal A}_{+-} & \equiv & {\cal A}(B_{d} \to {\ks}^{+} K^{*-})\;, \label{eq:amp-def-1}\qquad
{\cal A}_{-+}  \equiv {\cal A}(B_{d} \to {\ks}^{-} K^{*+})\;, \label{eq:amp-def-2} \non
\bar{{\cal A}}_{+-} &\equiv& {\cal A}(\bar B_{d} \to {\ks}^{+} K^{*-})\;,\label{eq:amp-def-3} \qquad
\bar{{\cal A}}_{-+}  \equiv  {\cal A}(\bar B_{d} \to {\ks}^{-} K^{*+})\;; \label{eq:amp-def-4}
\eeq
and
\beq
 \lambda_{+-}={q_{_{B_{d}}}\over p_{_{B_{d}}}}\,{\bar {\cal A}_{+-}\over {\cal A}_{+-}}\;, \qquad
 \lambda_{-+}={q_{_{B_{d}}}\over p_{_{B_{d}}}}\,{\bar {\cal A}_{-+}\over {\cal A}_{-+}}\;;
\eeq
where the vector $K^*$ meson is emitted by the $W$ boson in the case of ${\cal A}_{+-}$ and
$\bar {\cal A}_{-+}$, while it contains the spectator quark in the case of ${\cal A}_{-+}$
and $\bar {\cal A}_{+-}$.
Then one can get
\beq
 \acp^{\rm dir}+\Delta \acp^{\rm dir}&\equiv&{|\lambda_{+-}|^2-1\over |\lambda_{+-}|^2+1}=
{|\bar {\cal A}_{+-}|^2 -|{\cal A}_{+-}|^2\over |\bar {\cal A}_{+-}|^2+|{\cal A}_{+-}|^2}\;, \label{eq:acp-dir-p}
\non
 \acp^{\rm dir}-\Delta \acp^{\rm dir}&\equiv& {|\lambda_{-+}|^2-1\over |\lambda_{-+}|^2+1}={|\bar {\cal A}_{-+}|^2-|{\cal A}_{-+}|^2\over |\bar {\cal A}_{-+}|^2+|{\cal A}_{-+}|^2}\;, \label{eq:acp-dir-m}
\eeq
and
\beq
\acp^{\rm mix}+\Delta \acp^{\rm mix} &\equiv& {2\,{\rm Im}\lambda_{+-}\over |\lambda_{+-}|^2+1}={2\,{\rm Im}(e^{-2i\beta_{d}}\bar {\cal A}_{+-} {\cal A}_{+-}^*)\over
 |{\cal A}_{+-}|^2+|\bar {\cal A}_{+-}|^2}, \label{eq:acp-mix-p}\non
\acp^{\rm mix}-\Delta \acp^{\rm mix} &\equiv& {2\,{\rm Im}\lambda_{-+}\over |\lambda_{-+}|^2+1}={2\,{\rm Im}(e^{-2i\beta_{d}}\bar {\cal A}_{-+} {\cal A}_{-+}^*)\over
 |{\cal A}_{-+}|^2+|\bar {\cal A}_{-+}|^2}.\label{eq:acp-mix-m}
\eeq

Owing to the fact that $B_{d} \to {\ks}^{\pm} K^{*\mp}$ is not a CP eigenstate,
one must also consider the time- and flavor-integrated charge asymmetry,
\beq
 {\cal A}_{K_0^* K^*} &\equiv& {|{\cal A}_{+-}|^2+|\bar {\cal A}_{+-}|^2
 -|{\cal A}_{-+}|^2-|\bar {\cal A}_{-+}|^2\over |{\cal A}_{+-}|^2
 +|\bar {\cal A}_{+-}|^2+|{\cal A}_{-+}|^2+|\bar {\cal A}_{-+}|^2}\;, \label{eq:acp-tf}
\eeq
as another source of possible direct CP-violating asymmetry. Then, by transforming
the experimentally motivated direct CP parameters ${\cal A}_{K_0^*K^*}$ and
$\acp^{\rm dir}$ into the physically motivated choices, one can obtain the direct
CP asymmetries for $B_{d} \to {\ks}^+ K^{*-}$ and ${\ks}^- K^{*+}$ modes as the following,
\beq
 A_{{\ks}^+ K^{*-}} &\equiv&{\Gamma(\bar{B}_{d} \to {\ks}^{-}K^{*+})-\Gamma(B_{d}\to
 {\ks}^{+}K^{*-})\over  \Gamma(\bar{B}_{d} \to {\ks}^{-}K^{*+})+\Gamma(B_{d}\to
 {\ks}^{+}K^{*-})}= {|\kappa^{+-}|^2-1\over |\kappa^{+-}|^2+1}= - {{\cal A}_{K_0^*K^*}-\acp^{\rm dir}-{\cal A}_{K_0^*K^*}\Delta \acp^{\rm dir}\over 1-\Delta
 \acp^{\rm dir}-{\cal A}_{K_0^*K^*} \acp^{\rm dir}}\;, \label{eq:acp-pm}   \\
 A_{{\ks}^- K^{*+}} &\equiv& {\Gamma( \bar{B}_{d}\to {\ks}^+K^{*-})-\Gamma(B_{d}\to
 {\ks}^{-}K^{*+})\over \Gamma( \bar{B}_{d}\to {\ks}^+K^{*-})-\Gamma(B_{d}\to
 {\ks}^{-}K^{*+})}= {|\kappa^{-+}|^2-1\over |\kappa^{-+}|^2+1}= {{\cal A}_{K_0^*K^*}+\acp^{\rm dir}+{\cal A}_{K_0^*K^*}\Delta \acp^{\rm dir} \over 1+\Delta
 \acp^{\rm dir}+{\cal A}_{K_0^*K^*} \acp^{\rm dir}}\;; \label{eq:acp-mp}
\eeq
where
\beq
 \kappa^{+-}={q_{_{B_{d}}}\over p_{_{B_{d}}}}\,{\bar {\cal A}_{-+}\over {\cal A}_{+-}}\;, \qquad
  \kappa^{-+}={q_{_{B_{d}}}\over p_{_{B_{d}}}}\,{\bar {\cal A}_{+-}\over {\cal A}_{-+}}\;.
\eeq
Note that the difference among Eq.~(\ref{eq:acp-pm}) in this paper, Eq.~(161) in Ref.~\cite{CKMfitter}, and
Eq.~(5.20) in Ref.~\cite{Cheng09:b2m2m3} 
is just induced from the minus sign in the definition of direct CP-violating asymmetry, i.e.,
$\acp^{\rm dir}$, in
Eq.~(\ref{eq:acp-def}). The CP-violating parameters for the $B_s \to {\ks}^+ K^{*-}$
and $K^{*+} {\ks}^-$ decays can be similarly defined.

Based on the discussions on the CP violations in the above sector,
we can present the numerical results of the considered channels in both
scenarios for the CP-violating asymmetries in the pQCD approach
are as follows,
\beq
{\cal A}_{{\ks}K^*} &=&  \left\{ \begin{array}{ll}
\hspace{0.28cm}45.4^{+1.3}_{-1.7}(\omega_{b})
^{+0.1}_{-0.2}(\bar f_{K_0^*})
^{+0.0}_{-0.1}(f^{(T)}_{K^*})
^{+0.7}_{-1.1}(B_{i})
^{+0.8}_{-1.3}(a_{i})
^{+1.3}_{-2.2}(V_i)
^{+1.7}_{-2.5}(a_t)   \% &
\hspace{0.1cm} ({\rm S1}) \\
\hspace{0.28cm}46.1^{+1.1}_{-0.9}(\omega_{b})
^{+0.4}_{-0.4}(\bar f_{K_0^*})
^{+0.5}_{-0.2}(f^{(T)}_{K^*})
^{+2.6}_{-3.9}(B_{i})
^{+0.7}_{-1.0}(a_{i})
^{+1.4}_{-2.1}(V_i)
^{+4.2}_{-3.8}(a_t)   \% &
\hspace{0.1cm} ({\rm S2}) \\ \end{array} \right.   \label{eq:acp-tf-pm-bd}  \;,
\eeq
 \beq
\acp^{\rm dir} &=&  \left\{ \begin{array}{ll}
\hspace{0.18cm}-5.7^{+0.3}_{-0.4}(\omega_{b})
^{+0.1}_{-0.0}(\bar f_{K_0^*})
^{+0.7}_{-0.2}(f^{(T)}_{K^*})
^{+0.5}_{-0.7}(B_{i})
^{+2.9}_{-1.4}(a_{i})
^{+0.3}_{-0.3}(V_i)
^{+0.3}_{-0.3}(a_t)   \% &
\hspace{0.1cm} ({\rm S1}) \\
-16.0^{+0.1}_{-0.1}(\omega_{b})
^{+0.3}_{-0.3}(\bar f_{K_0^*})
^{+0.7}_{-0.8}(f^{(T)}_{K^*})
^{+4.2}_{-4.9}(B_{i})
^{+3.4}_{-2.9}(a_{i})
^{+1.0}_{-0.5}(V_i)
^{+1.3}_{-1.8}(a_t)   \% &
\hspace{0.1cm} ({\rm S2}) \\ \end{array} \right.   \label{eq:acp-dir-pm-bd}  \;,\\
\Delta \acp^{\rm dir} &=&  \left\{ \begin{array}{ll}
\hspace{0.28cm}86.4^{+0.8}_{-0.7}(\omega_{b})
^{+0.1}_{-0.2}(\bar f_{K_0^*})
^{+0.1}_{-0.1}(f^{(T)}_{K^*})
^{+0.2}_{-0.3}(B_{i})
^{+1.4}_{-1.9}(a_{i})
^{+1.4}_{-1.1}(V_i)
^{+1.7}_{-1.7}(a_t)   \% &
\hspace{0.1cm} ({\rm S1}) \\
\hspace{0.28cm}74.0^{+1.1}_{-1.1}(\omega_{b})
^{+0.5}_{-0.8}(\bar f_{K_0^*})
^{+0.0}_{-0.1}(f^{(T)}_{K^*})
^{+1.9}_{-4.8}(B_{i})
^{+0.4}_{-1.0}(a_{i})
^{+0.8}_{-1.2}(V_i)
^{+0.0}_{-0.1}(a_t)   \% &
\hspace{0.1cm} ({\rm S2}) \\ \end{array} \right.   \label{eq:acp-dird-pm-bd}  \;,
\eeq
 \beq
\acp^{\rm mix} &=&  \left\{ \begin{array}{ll}
-14.2^{+0.3}_{-1.2}(\omega_{b})
^{+0.1}_{-0.0}(\bar f_{K_0^*})
^{+0.2}_{-0.4}(f^{(T)}_{K^*})
^{+1.4}_{-1.7}(B_{i})
^{+0.4}_{-0.4}(a_{i})
^{+2.1}_{-0.9}(V_i)
^{+2.7}_{-4.0}(a_t)   \% &
\hspace{0.1cm} ({\rm S1}) \\
\hspace{0.3cm}10.0^{+0.8}_{-0.8}(\omega_{b})
^{+0.6}_{-0.3}(\bar f_{K_0^*})
^{+0.9}_{-0.6}(f^{(T)}_{K^*})
^{+2.2}_{-4.9}(B_{i})
^{+2.8}_{-3.0}(a_{i})
^{+6.2}_{-4.1}(V_i)
^{+0.1}_{-1.1}(a_t)   \% &
\hspace{0.1cm} ({\rm S2}) \\ \end{array} \right.   \label{eq:acp-mix-pm-bd}  \;,\\
\Delta\acp^{\rm mix} &=&  \left\{ \begin{array}{ll}
\hspace{0.18cm}-1.8^{+0.7}_{-0.5}(\omega_{b})
^{+0.2}_{-0.4}(\bar f_{K_0^*})
^{+0.1}_{-0.0}(f^{(T)}_{K^*})
^{+1.6}_{-1.7}(B_{i})
^{+2.0}_{-1.2}(a_{i})
^{+2.1}_{-1.6}(V_i)
^{+1.0}_{-0.3}(a_t)  \% &
\hspace{0.1cm} ({\rm S1}) \\
-28.1^{+1.8}_{-2.0}(\omega_{b})
^{+0.3}_{-0.5}(\bar f_{K_0^*})
^{+1.5}_{-1.5}(f^{(T)}_{K^*})
^{+18.9}_{-7.8}(B_{i})
^{+8.2}_{-7.7}(a_{i})
^{+0.8}_{-0.4}(V_i)
^{+0.5}_{-0.0}(a_t)   \% &
\hspace{0.1cm} ({\rm S2}) \\ \end{array} \right.   \label{eq:acp-mixd-pm-bd}  \;,
\eeq
 \beq
A_{{\ks}^+K^{*-}} &=&  \left\{ \begin{array}{ll}
-73.2^{+1.3}_{-2.5}(\omega_{b})
^{+0.5}_{-0.0}(\bar f_{K_0^*})
^{+2.8}_{-1.1}(f^{(T)}_{K^*})
^{+1.9}_{-2.7}(B_{i})
^{+14.6}_{-8.3}(a_{i})
^{+2.1}_{-3.0}(V_i)
^{+2.6}_{-3.2}(a_t)  \% &
\hspace{0.1cm} ({\rm S1}) \\
-83.9^{+0.7}_{-0.7}(\omega_{b})
^{+0.6}_{-0.2}(\bar f_{K_0^*})
^{+1.5}_{-1.4}(f^{(T)}_{K^*})
^{+5.7}_{-2.9}(B_{i})
^{+7.9}_{-5.5}(a_{i})
^{+3.5}_{-1.6}(V_i)
^{+4.0}_{-4.4}(a_t)  \% &
\hspace{0.1cm} ({\rm S2}) \\ \end{array} \right.   \label{eq:acp-dir-k0spksm-bd}  \;,\\
A_{{\ks}^-K^{*+}} &=&  \left\{ \begin{array}{ll}
\hspace{0.28cm}43.0^{+1.4}_{-2.0}(\omega_{b})
^{+0.1}_{-0.2}(\bar f_{K_0^*})
^{+0.1}_{-0.4}(f^{(T)}_{K^*})
^{+0.8}_{-1.4}(B_{i})
^{+1.4}_{-1.9}(a_{i})
^{+1.5}_{-2.5}(V_i)
^{+1.8}_{-2.7}(a_t)   \% &
\hspace{0.1cm} ({\rm S1}) \\
\hspace{0.28cm}38.5^{+1.1}_{-0.8}(\omega_{b})
^{+0.6}_{-0.6}(\bar f_{K_0^*})
^{+0.2}_{-0.0}(f^{(T)}_{K^*})
^{+4.6}_{-6.7}(B_{i})
^{+1.1}_{-1.1}(a_{i})
^{+1.5}_{-1.9}(V_i)
^{+3.7}_{-3.4}(a_t)  \% &
\hspace{0.1cm} ({\rm S2}) \\ \end{array} \right.   \label{eq:acp-dir-k0smksp-bd}  \;,
\eeq
for $B_{d} \to {\ks}^+ K^{*-}$ and ${\ks}^- K^{*+}$ decays,
and
\beq
{\cal A}_{{\ks}K^*} &=&  \left\{ \begin{array}{ll}
\hspace{0.45cm}0.5^{+5.2}_{-5.0}(\omega_{bs})
^{+0.6}_{-0.5}(\bar f_{K_0^*})
^{+2.5}_{-2.3}(f^{(T)}_{K^*})
^{+1.8}_{-2.2}(B_{i})
^{+8.7}_{-7.5}(a_{i})
^{+0.0}_{-0.0}(V_i)
^{+0.8}_{-2.2}(a_t)   \% &
\hspace{0.1cm} ({\rm S1}) \\
-20.3^{+2.5}_{-1.7}(\omega_{bs})
^{+0.5}_{-0.4}(\bar f_{K_0^*})
^{+0.1}_{-0.1}(f^{(T)}_{K^*})
^{+1.7}_{-0.4}(B_{i})
^{+3.0}_{-2.0}(a_{i})
^{+0.6}_{-0.8}(V_i)
^{+0.6}_{-0.4}(a_t)   \% &
\hspace{0.1cm} ({\rm S2}) \\ \end{array} \right.   \label{eq:acp-tf-pm-bs}  \;,
\eeq
 \beq
\acp^{\rm dir} &=&  \left\{ \begin{array}{ll}
\hspace{0.45cm}3.2^{+4.2}_{-7.4}(\omega_{bs})
^{+0.5}_{-0.5}(\bar f_{K_0^*})
^{+2.8}_{-2.2}(f^{(T)}_{K^*})
^{+2.9}_{-3.6}(B_{i})
^{+12.7}_{-14.2}(a_{i})
^{+0.1}_{-0.1}(V_i)
^{+0.6}_{-2.5}(a_t)   \% &
\hspace{0.1cm} ({\rm S1}) \\
-27.2^{+4.1}_{-2.9}(\omega_{bs})
^{+0.2}_{-0.1}(\bar f_{K_0^*})
^{+1.3}_{-1.1}(f^{(T)}_{K^*})
^{+1.8}_{-0.2}(B_{i})
^{+4.6}_{-2.7}(a_{i})
^{+0.9}_{-1.1}(V_i)
^{+0.5}_{-0.0}(a_t)   \% &
\hspace{0.1cm} ({\rm S2}) \\ \end{array} \right.   \label{eq:acp-dir-pm-bs}  \;,\\
\Delta \acp^{\rm dir} &=&  \left\{ \begin{array}{ll}
-11.5^{+21.7}_{-26.8}(\omega_{bs})
^{+0.6}_{-0.3}(\bar f_{K_0^*})
^{+6.1}_{-6.1}(f^{(T)}_{K^*})
^{+14.8}_{-13.4}(B_{i})
^{+18.1}_{-15.7}(a_{i})
^{+2.4}_{-4.1}(V_i)
^{+12.9}_{-13.6}(a_t)   \% &
\hspace{0.1cm} ({\rm S1}) \\
-45.4^{+5.0}_{-4.0}(\omega_{bs})
^{+0.1}_{-0.0}(\bar f_{K_0^*})
^{+3.8}_{-3.5}(f^{(T)}_{K^*})
^{+2.8}_{-5.5}(B_{i})
^{+4.5}_{-2.2}(a_{i})
^{+1.4}_{-2.4}(V_i)
^{+4.2}_{-3.4}(a_t)   \% &
\hspace{0.1cm} ({\rm S2}) \\ \end{array} \right.   \label{eq:acp-dird-pm-bs}  \;,
\eeq
 \beq
\acp^{\rm mix} &=&  \left\{ \begin{array}{ll}
\hspace{0.28cm}36.5^{+11.5}_{-17.5}(\omega_{bs})
^{+0.5}_{-0.6}(\bar f_{K_0^*})
^{+3.5}_{-3.6}(f^{(T)}_{K^*})
^{+6.9}_{-10.5}(B_{i})
^{+7.1}_{-9.1}(a_{i})
^{+1.4}_{-1.2}(V_i)
^{+1.6}_{-4.3}(a_t)   \% &
\hspace{0.1cm} ({\rm S1}) \\
-27.4^{+7.9}_{-5.3}(\omega_{bs})
^{+0.2}_{-0.2}(\bar f_{K_0^*})
^{+0.5}_{-0.3}(f^{(T)}_{K^*})
^{+12.6}_{-5.1}(B_{i})
^{+3.9}_{-3.2}(a_{i})
^{+1.0}_{-0.8}(V_i)
^{+2.9}_{-2.5}(a_t)   \% &
\hspace{0.1cm} ({\rm S2}) \\ \end{array} \right.   \label{eq:acp-mix-pm-bs}  \;,\\
\Delta\acp^{\rm mix} &=&  \left\{ \begin{array}{ll}
-62.3^{+16.6}_{-14.8}(\omega_{bs})
^{+0.6}_{-0.8}(\bar f_{K_0^*})
^{+5.2}_{-4.6}(f^{(T)}_{K^*})
^{+10.3}_{-9.6}(B_{i})
^{+9.0}_{-5.4}(a_{i})
^{+2.0}_{-1.2}(V_i)
^{+0.6}_{-0.2}(a_t)   \% &
\hspace{0.1cm} ({\rm S1}) \\
-35.3^{+2.7}_{-1.2}(\omega_{bs})
^{+0.5}_{-0.4}(\bar f_{K_0^*})
^{+5.6}_{-5.4}(f^{(T)}_{K^*})
^{+23.4}_{-14.7}(B_{i})
^{+8.8}_{-9.2}(a_{i})
^{+3.1}_{-1.5}(V_i)
^{+7.4}_{-5.3}(a_t)   \% &
\hspace{0.1cm} ({\rm S2}) \\ \end{array} \right.   \label{eq:acp-mixd-pm-bs}  \;,
\eeq
 \beq
A_{{\ks}^+K^{*-}} &=&  \left\{ \begin{array}{ll}
\hspace{0.12cm}2.4^{+0.2}_{-0.9}(\omega_{bs})
^{+0.9}_{-1.1}(\bar f_{K_0^*})
^{+0.3}_{-0.2}(f^{(T)}_{K^*})
^{+0.9}_{-1.4}(B_{i})
^{+5.6}_{-4.5}(a_{i})
^{+0.1}_{-0.1}(V_i)
^{+0.1}_{-0.1}(a_t)   \% &
\hspace{0.1cm} ({\rm S1}) \\
\hspace{0.12cm}1.7^{+0.3}_{-0.4}(\omega_{bs})
^{+0.6}_{-0.6}(\bar f_{K_0^*})
^{+0.3}_{-0.2}(f^{(T)}_{K^*})
^{+1.2}_{-1.1}(B_{i})
^{+0.4}_{-0.4}(a_{i})
^{+0.1}_{-0.0}(V_i)
^{+0.1}_{-0.1}(a_t)   \% &
\hspace{0.1cm} ({\rm S2}) \\ \end{array} \right.   \label{eq:acp-dir-k0spksm-bs}  \;,\\
A_{{\ks}^-K^{*+}} &=&  \left\{ \begin{array}{ll}
\hspace{0.45cm}4.1^{+8.2}_{-15.4}(\omega_{bs})
^{+0.1}_{-0.0}(\bar f_{K_0^*})
^{+6.1}_{-4.8}(f^{(T)}_{K^*})
^{+5.9}_{-6.0}(B_{i})
^{+19.6}_{-25.5}(a_{i})
^{+0.3}_{-0.1}(V_i)
^{+1.0}_{-4.8}(a_t)   \% &
\hspace{0.1cm} ({\rm S1}) \\
-63.6^{+10.7}_{-8.4}(\omega_{bs})
^{+0.1}_{-0.0}(\bar f_{K_0^*})
^{+4.4}_{-4.5}(f^{(T)}_{K^*})
^{+4.0}_{-2.8}(B_{i})
^{+11.4}_{-6.7}(a_{i})
^{+1.9}_{-3.2}(V_i)
^{+3.2}_{-2.1}(a_t)   \% &
\hspace{0.1cm} ({\rm S2}) \\ \end{array} \right.   \label{eq:acp-dir-k0smksp-bs}  \;,
\eeq
for $B_{s} \to {\ks}^+ K^{*-}$ and ${\ks}^- K^{*+}$ decays.

For the direct CP asymmetries in
the pure annihilation $B_d \to {\ks}^+ K^{*-}$ and $B_d \to K^{*+} {\ks}^-$ decays as defined in Eqs.~(\ref{eq:acp-pm}) and
(\ref{eq:acp-mp})
for example, one
can find from the numerical results shown in Eqs.~(\ref{eq:acp-dir-k0spksm-bd}) and (\ref{eq:acp-dir-k0smksp-bd})
that their signs and magnitudes are rather different in these two modes within theoretical errors. In the former mode,
the direct CP violation is about -73\% in S1 and -84\% in S2; while in the latter one, the direct CP asymmetry
is 43\% in S1 and 39\% in S2. It is clear to find that
the magnitudes of direct CP-violating asymmetries predicted in the pQCD approach for these two modes
in both scenarios are much large, which can be tested at the ongoing LHC and
forthcoming Super-$B$ experiments, by combining the large branching ratios($\sim {\cal O}(10^{-6})$). Furthermore, once the predictions
on the physical quantities in the pQCD approach could be confirmed at the predicted
level by the precision experimental measurements in the future, which can also provide indirect evidences
for the important but
controversial issue on the evaluation of annihilation
contributions at leading power: almost real with tiny strong phase in soft collinear effective theory
or almost imaginary with large strong phase in the pQCD approach?

\section{Summary} \label{sec:summary}

In this work, we studied the charmless hadronic $B_{u/d/s} \to \ks K^*$ decays by employing the pQCD
approach based on the framework of $k_T$ factorization theorem. By regarding the scalar $\ks$ as the
conventional $q\bar q$ meson, then with the help of the light-cone distribution amplitude of $\ks$ up to twist-3
in two scenarios, we explored the physical observables such as branching ratios and CP-violating asymmetries
of the considered channels. It is worth mentioning that, in this paper, as the first estimates to the physical
observables of $B \to \ks K^*$ decays, only the perturbatively short distance contributions at leading order
are investigated. We do not consider the possible long-distance
contributions, such as the rescattering effects, although they should be present, and they
may be large and affect the theoretical predictions. It is beyond
the scope of this work and expected to be studied in the future.

From the numerical evaluations and phenomenological analysis in the pQCD approach, we found the following
results:

\begin{itemize}
\item
The considered $B_u \to {\ks}^+ \overline{K}^{*0}$ and $K^{*+} {\ksb}^0$ decays exhibit large
branching ratios($\sim {\cal O}(10^{-6})$) and large direct CP-violating asymmetries in S2, which
are clearly measurable in B factories and LHC experiments and are helpful to better understand the
QCD behavior of the scalar $\ks$ in turn.

\item
In the considered modes, only the preliminary upper limit on $Br(B_d \to {\ks}^0 \overline{K}^{*0})$ has been
reported by Belle collaboration. The predicted results agree basically with this upper limit and will be
tested by the more precision measurements in the future.

\item
Most of the considered decays are affected significantly by the involved weak annihilation contributions.
The predictions on large
branching ratios and large direct CP violations of the pure annihilation processes $B_d \to {\ks}^+ K^{*-}$ and $K^{*+} {\ks}^-$ can be measured in the ongoing LHC experiments and forthcoming Super-B
factory, which will provide more evidences to help understand the annihilation contributions in
B physics.

\item
The $B_{d/s} \to {\ks}^0 \overline{K}^{*0}$ and $K^{*0} {\ksb}^0$ decays can be viewed as a good platform to test
the exotic new physics beyond the SM if the obviously nonzero direct CP violations could be observed.

\item
Generally speaking, the pQCD predictions for the considered decays still suffer from
large theoretical errors induced by the uncertainties of the input parameters, e.g., mesonic
decay constants, Gegenbauer moments in the universal distribution amplitudes, etc., which
are expected to be constrained by the more and more precision data.

\end{itemize}


\begin{acknowledgments}
X.~Liu thanks Y.M.~Wang for his discussions and comments.
This work is supported by the National Natural Science
Foundation of China under Grants Nos.~11205072 and~11235005,
and by a project funded by the Priority Academic Program Development
of Jiangsu Higher Education Institutions (PAPD),
and by the Research Fund of Jiangsu Normal University under Grant No.~11XLR38.
\end{acknowledgments}


\end{document}